\documentclass[lettersize,journal]{IEEEtran}

\usepackage{url}
\usepackage{graphicx}
\usepackage{subfigure}
\usepackage{amsmath} 
\usepackage{bm}
\usepackage{amsfonts,amssymb} 
\usepackage{pifont}
\usepackage{booktabs}
\usepackage{bbding}
\usepackage{makecell}
\usepackage{xcolor}
\usepackage{wasysym}
\usepackage{multirow}
\usepackage[numbers,sort&compress]{natbib}
\usepackage[pagebackref=false,breaklinks=true,colorlinks,bookmarks=false,urlcolor=magenta]{hyperref}
\usepackage[linesnumbered,ruled,vlined]{algorithm2e}
\definecolor{darkblue}{rgb}{0, 0, 0.5}
\hypersetup{
  colorlinks=true,
  linkcolor=blue,
  citecolor=blue,
  urlcolor=black
}

\begin{document}

\title{FortisAVQA and MAVEN: a Benchmark Dataset and Debiasing Framework for Robust Multimodal Reasoning}

\author{Jie Ma,~\IEEEmembership{Member,~IEEE,}
        Zhitao Gao\textsuperscript{\textdagger},
        Qi Chai\textsuperscript{\textdagger}, 
        Jun Liu,~\IEEEmembership{Senior Member,~IEEE,}
        Pinghui Wang, ~\IEEEmembership{Senior Member,~IEEE,}
        Jing Tao,
        Zhou Su
        % <-this % stops a space
\IEEEcompsocitemizethanks{
\IEEEcompsocthanksitem This work was supported in part by the National Natural Science Foundation of China (62306229, 62137002, U22B2019, 62477037, 62450005, 62437002, 62293553), the Natural Science Basic Research Program of Shaanxi (2023-JC-YB-593), the Youth Innovation Team of Shaanxi Universities ``Multi-modal Data Mining and Fusion'', the Shaanxi Undergraduate and Higher Education Teaching Reform Research Program (23BY195), the Youth Talent Support Program of Shaanxi Science and Technology Association (20240113), and the China Postdoctoral Science Foundation (2024M752585).
\IEEEcompsocthanksitem Jie Ma, Pinghui Wang, Jing Tao and Zhou Su are with the  Ministry of Education of Key Laboratory for Intelligent Networks and Network Security, School of Cyber Science and Engineering, Xi'an Jiaotong University, Xi'an, Shaanxi 710049, China. 
\IEEEcompsocthanksitem Zhitao Gao and Jun Liu are with the Shannxi Provincial Key Laboratory of Big Data Knowledge Engineering, School of Computer Science and Technology, Xi'an Jiaotong University, Xi'an, Shaanxi 710049, China.
\IEEEcompsocthanksitem Qi Chai is with the Information Hub, Hong Kong University of Science and Technology (Guangzhou), Guangzhou, Guangdong, 510000, China.
%\IEEEcompsocthanksitem Jun Liu and Pinghui Wang are the corresponding authors (liukeen@xjtu.edu.cn, phwang@mail.xjtu.edu.cn).
\IEEEcompsocthanksitem \textdagger denotes equal contribution.
% <-this % stops a space
}
}

\markboth{Under Review}%,~Vol.~14, No.~8, August~2015}%
{Ma \MakeLowercase{\textit{et al.}}: FortisAVQA and MAVEN: a Benchmark Dataset and Debiasing Framework for Robust Multimodal Reasoning}


\IEEEtitleabstractindextext{
\begin{abstract}
Audio-Visual Question Answering (AVQA) is a challenging multimodal reasoning task requiring intelligent systems to answer natural language queries based on paired audio-video inputs accurately. However, existing AVQA approaches often suffer from overfitting to dataset biases, leading to poor robustness. Moreover, current datasets may not effectively diagnose these methods. To address these challenges, we first introduce a novel dataset, FortisAVQA, constructed in two stages: (1) rephrasing questions in the test split of the public MUSIC-AVQA dataset and (2) introducing distribution shifts across questions. The first stage expands the test space with greater diversity, while the second enables a refined robustness evaluation across rare, frequent, and overall question distributions. Second, we introduce a robust Multimodal Audio-Visual Epistemic Network (MAVEN) that leverages a multifaceted cycle collaborative debiasing strategy to mitigate bias learning.  Experimental results demonstrate that our architecture achieves state-of-the-art performance on FortisAVQA, with a notable improvement of 7.81\%. Extensive ablation studies on both datasets validate the effectiveness of our debiasing components. Additionally, our evaluation reveals the limited robustness of existing multimodal QA methods. We also verify the plug-and-play capability of our strategy by integrating it with various baseline models across both datasets. Our dataset and code are available at \url{https://github.com/reml-group/fortisavqa}.
\end{abstract}

}
\maketitle
\IEEEdisplaynontitleabstractindextext
%\IEEEpeerreviewmaketitle

\section{Introduction}
\label{sec:intro}
\IEEEPARstart{H}{umans} possess the extraordinary capacity to seamlessly integrate auditory and visual cues, effectively establishing a cohesive relationship between visual and auditory stimuli \cite{lin2023vision,ma2024diagram,ma2022weakly}. Audio-Visual Question Answering (AVQA) \cite{alamri2019audio, yang2022avqa,li2022learning,yun2021pano} seeks to enable intelligent systems to acquire this capability and produce answers based on provided natural language questions. It requires the system to learn high-order interaction representations of the concepts encompassed with audio, video, and language modalities. As is known to us \cite{wen2021debiased,vatsa2024adventures,hall2024visogender}, the high-level reasoning ability of the system mainly relies on large-scale data that does not contain harmful biases or statistical regularities.

\begin{figure}[tbp]
    \centering  %图片全局居中
    \includegraphics[width=\linewidth]{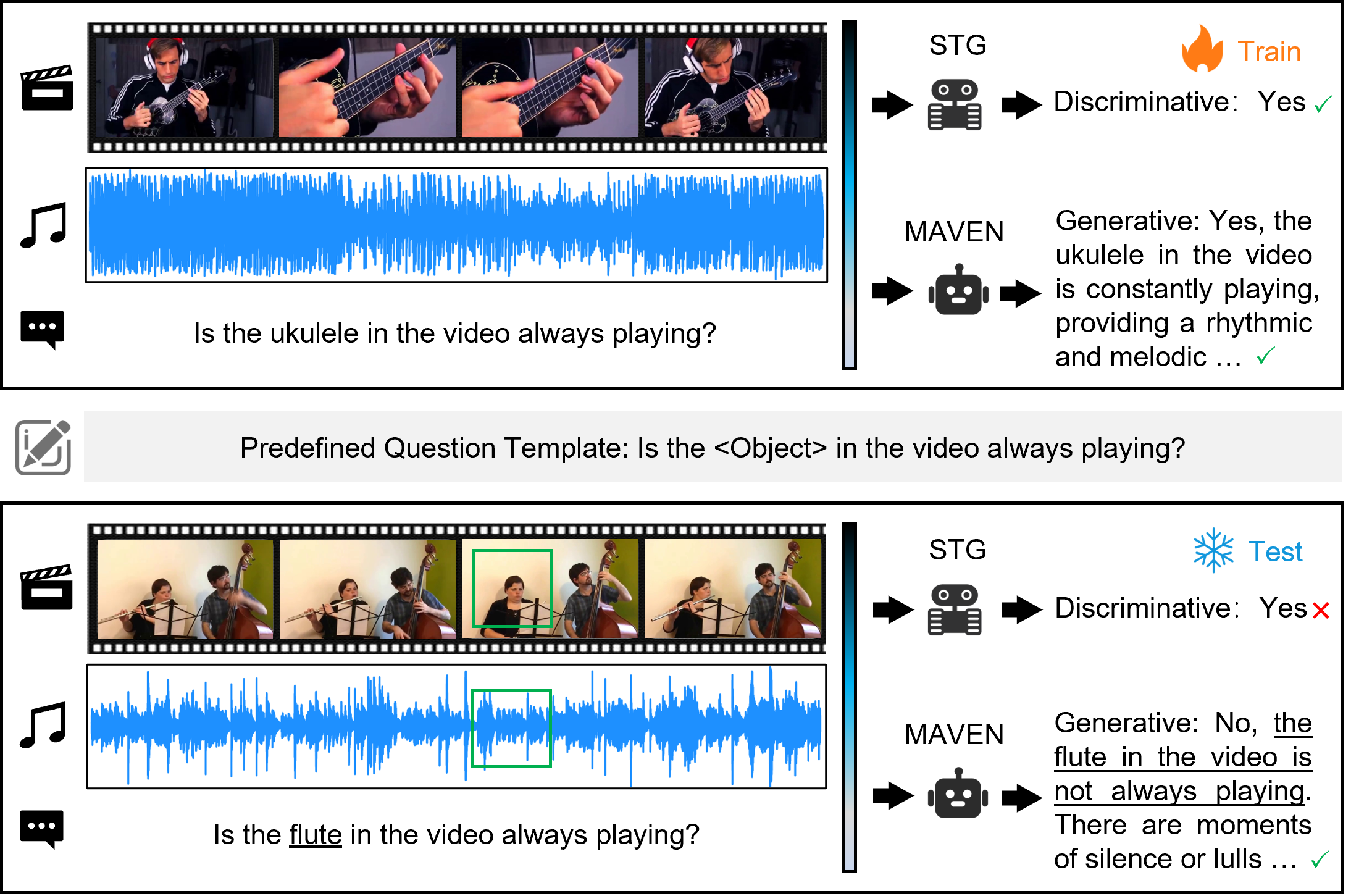}
    \caption{An example illustrating the existing AVQA dataset construction and the comparison between STG and MAVEN. The question in current AVQA datasets is generated by a limited set of predefined templates, which may not be in line with the real-world scenario. Our findings indicate that existing methods such as STG \cite{li2022learning} are not robust, which may be attributed to excessive bias learning, such as memorizing statistical regularities between critical question words and answers.}
    \label{fig:avqa-example}
\end{figure}

However, completely avoiding the negative bias in datasets seems challenging \cite{torralba2011} due to the inherent skewness in real-world data distributions. Previous studies \cite{li2023multi,ravichander2023and} in visual and extractive QA have investigated the bias from the perspective of answer distribution changing \cite{agrawal2018don,kervadec2021roses} and language (question) bias elimination \cite{niu2021introspective,xue2024integrating}. Drawing inspiration from these studies, we propose several open research questions for the AVQA task, focusing on dataset evaluations and model design perspectives.

\textbf{Question 1: Have existing datasets comprehensively measured model robustness?} The questions in the current AVQA dataset \cite{yun2021pano,liu2024tackling,yang2022avqa} are generated by a limited set of predefined templates, such as the 33 templates in the MUSIC-AVQA dataset \cite{li2022learning}. Fig. \ref{fig:avqa-example} shows the samples in the training and test split, which are produced using a predefined template. The observed difference mainly stems from a single word, leading to a limited vocabulary size of only 93 words. This has the potential to deviate from real-world scenarios. Moreover, current datasets cannot reflect the performance on rare or less common samples, which is an important indicator for evaluating model robustness \cite{zhangZX24,kervadec2021roses}. 

\textbf{Question 2: Have existing methods overcome the data bias?} We observed that existing methods \cite{yun2021pano, lin2023vision, girdhar2023, zhangLB23} like STG \cite{li2022learning}, exhibit brittleness when handling questions with rare answers during the test stage. This vulnerability may stem from the tendency of models to memorize statistical regularities between key question words and answers, such as the association between ``Is," ``Playing," and ``Yes". Notably, experimental results \cite{li2022learning} indicate that STG achieves an accuracy of 54.09\% on the test split of MUSIC-AVQA using only questions, despite AVQA being a multimodal task that integrates audio, video, and text modalities.

In this paper, we present the development of a novel dataset called FortisAVQA, which aims to address the first question precisely. The dataset complements MUSIC-AVQA \cite{li2022learning} and provides a more refined diagnostic for AVQA methods. \textit{To preserve the inherent bias, we maintain the original training and validation splits of the MUSIC-AVQA dataset.} In contrast, we employ a human-machine collaboration mechanism to rephrase the question in the test split. This ensures diverse and natural question forms while remarkably expanding the number of questions from \textbf{9,129} to \textbf{211,572}. We introduce a distribution shift based on the answer distributions of specific question types. This allows us to measure performance on both frequent (in-distribution) and rare (out-of-distribution) data simultaneously.  

To tackle the second question, we propose a robust framework that applies a \textit{Multifaceted Cycle Collaborative Debiasing} (MCCD) strategy. Specifically, the strategy introduces a novel optimization objective, which enlarges the distribution difference between unimodal (question, audio, and video) and multimodal logit. By doing so, our model becomes less prone to learning biases from individual modalities. Intuitively, we cannot choose the accurate answer based on only one modality. Hence, MCCD employs cycle guidance to constrain the logit distribution of each modality, thereby promoting the similarity of unimodal logit distribution. The experimental results demonstrate that our framework yields significant improvements on both datasets, with a particularly notable enhancement of 7.81\% observed on the FortisAVQA dataset.

This work represents an improved and extended version of our previously published paper in NeurIPS 2024 \cite{ma2024look}. In comparison to our prior research, this paper introduces several key advancements. First, we present a novel mechanism for introducing distribution shifts that automatically and reasonably separate head and tail questions, drawing inspiration from conformal prediction \cite{oliveira2024split}. Second, we propose a Multimodal Audio-Visual Epistemic Network (MAVEN) utilizing MCCD for robust generation. To our knowledge, this is the first bias-mitigating exploration in \textit{generative} AVQA architectures. Third, we refine the technique for enlarging the distribution difference between unimodal and multimodal logit within MCCD, transforming distance measurement into a KL divergence evaluation. These contributions significantly enhance both the interactivity of AVQA models and the rationality of evaluations, providing insights for future research on unbiased generative architectures.

In summary, our contributions are fourfold. 
\begin{itemize}
    \item We introduce FortisAVQA, a novel dataset, along with a comprehensive set of evaluation metrics. This enables a rigorous assessment of the reasoning capabilities of AVQA models, providing insights into their generalization performance across both in-distribution and out-of-distribution scenarios.
    \item We present an AVQA architecture MAVEN that incorporates the MCCD strategy to overcome training biases and perform robust generation. To the best of our knowledge, this is the first work to systematically explore biases in the AVQA task from dataset evaluations as well as model designs.
    \item We conduct extensive experiments on MUSIC-AVQA and FortisAVQA to verify the effectiveness and superiority of MAVEN and debiasing strategy.
    \item We evaluate 12 recent multimodal QA methods on the proposed dataset and show their limited ability to generalize not only in-distribution scenarios but also in out-of-distribution situations.
\end{itemize}

\section{Related Work}
\subsection{Multimodality Learning}
Drawing inspiration from human cognition, AI researchers have been vigorously exploring deep learning frameworks to integrate multiple modalities \cite{lyu2023macaw}, such as audio, vision, and text. This growing interest is driven in part by the remarkable advancements in large language models (LLMs) \cite{guo2025deepseek,bai2023qwen} and Transformer architectures, which have offered valuable insights that support further research on multimodal learning.

A key research direction in multimodal learning centers on constructing joint embedding spaces that seamlessly integrate multiple modalities in an end-to-end fashion \cite{li2021align, radford2021learning, li2022blip}. Pioneering models such as CLIP \cite{radford2021learning} and ALIGN \cite{jia2021scaling} exemplify this paradigm by aligning visual and textual representations within a shared latent space, thereby enabling robust zero-shot recognition and retrieval capabilities. Additionally, models such as MLVQA \cite{ma2021multitask} and CoCa \cite{yu2022coca} leverage cross-attention mechanisms within Transformer architectures and are trained with multiple loss functions to enhance multimodal feature fusion. These advancements collectively contribute to more effective and generalizable multimodal representations, facilitating improved downstream performance across a wide range of tasks.

Another research direction seeks to combine pre-trained vision-only and language-only models, leveraging their complementary strengths to improve generalization and achieve remarkable zero-shot learning capabilities \cite{li2023blip,jian2024,wu2024building}. Hybrid losses, such as mask language modeling and image-text contrastive learning, are employed to optimize the embedding of queries. Such approaches highlight the potential of integrating modality-specific pre-training paradigms to enhance multimodal learning.

More recently, the community has turned its attention to instruction tuning for multimodal LLMs. This paradigm equips large models with the ability to follow human instructions across modalities, thereby broadening their applicability in real-world scenarios \cite{dai2023instructblip,zhuminigpt,cheng2024videollama,cheng2024videollama,fu2024vita}. A significant milestone in this direction is MultiInstruct, the first large-scale multimodal instruction-tuning benchmark dataset, covering a broad spectrum of multimodal tasks and domains \cite{xu2023multiinstruct}. Researchers have also explored synthetic data-driven instruction tuning \cite{zhang2024video} and parameter-efficient fine-tuning methods such as LoRA, enabling textual LLMs to handle up to six modalities with minimal computational overhead \cite{su2023pandagpt}. Despite recent advances, existing multimodal models remain limited by their inability to concurrently process audio, video, and text inputs, and they continue to be affected by inherent biases in the training data.

\subsection{Model Robustness Evaluation}
Existing QA datasets \cite{goyal2019making,hudson2019gqa,yang2022,yang2022avqa} suffer from biases, resulting in imprecise evaluations. In recent years, numerous studies have tackled this issue from various perspectives \cite{ma2023robust, kervadec2021, ramakrishnanAL18, zheng2023large}.

One avenue of research \cite{agrawal2018don,ko2020look,dancette2021beyond} reorganizes existing datasets, thereby making the distribution between training and testing splits significantly different or even reversed. The reorganized datasets reflect the performance in the out-of-distribution situation but lack measurement in the in-distribution scenario. To this end, GQA-OOD \cite{kervadec2021roses} introduces the distribution shift in both the validation and test splits to assess visual QA models in both scenarios simultaneously. Nevertheless, the number of questions in the GQA-OOD test split is only 2,796, which may not reflect the real generalization ability of visual QA models due to the presence of a limited number of testing samples \cite{harrell1996}. Inspired by the adversarial attack, another line of works \cite{sheng2021human,li2021adversarial} regard the dataset construction as a game played by two parties: a human annotator and a well-trained model.  Only samples generated by humans that successfully attack the model are incorporated into the dataset. In addition, there exists another line of work \cite{liu2024tackling} that complements videos and questions to obtain balanced training data.

Different from the mentioned works, FortisAVQA not only prioritizes question diversity but also considers the volume of test samples. This enhances the precision and comprehensiveness of robustness evaluation. Moreover, we recognize the formidable challenge of obtaining completely pure training data. As such, we opt to retain the inherent bias present in both the training and validation splits. Our primary objective is to inspire the community to enhance model robustness through the implementation of debiasing strategies rather than striving for balanced training data. Remarkably, to the best of our knowledge, our dataset is the first AVQA dataset explicitly designed for robustness evaluation.

\subsection{Bias Dependency Elimination}
A variety of debiasing QA methods \cite{xu2023counterfactual, TsirigotisMR0C23, esiobu2023robbie} have been proposed to overcome bias memorization. These methods can be divided into four classes \cite{ma2023robust}: ensemble learning, data augmentation, contrastive learning, and answer re-ranking.

Ensemble learning methods \cite{niu2021introspective,wen2021debiased,Cho0RK23,ma2023adaptive} typically leverage a combination of a bias learner and a vanilla QA model to comprehensively predict answers. Data augmentation methods \cite{abbasnejad2020counterfactual,chen2022rethinking,teney2021unshuffling} generate additional question-answer pairs to balance the data distribution. Based on the positive and negative sample generation, contrastive learning-based methods \cite{liang2020learning,zhu2021overcoming,si2022towards} strive to learn an embedding space where similar sample pairs are closely clustered while disparate ones are distinctly separated. Consequently, the vanilla QA method is optimized jointly through contrastive and QA losses. Answer re-ranking methods \cite{jing2020overcoming,shrestha2020negative,gat2020removing} primarily focus on reordering the answers predicted by the vanilla QA model to enhance context comprehension, such as vision grounding. 

To the best of our knowledge, COCA \cite{lao2023coca} is the only work to mitigate the bias learning in the AVQA task, which first employs causal regularization to intervene bias-irrelevant causal effects and then introspects predictions. Unlike the mentioned works, which only consider language biases, our method considers audio, vision, language biases, and their collaboration. The proposed MCCD strategy features plug-and-play capability, enhancing the debiasing potential of baseline methods.

%In recent years, pre-trained multimodal models \cite{lu2019vilbert,wu2019self,tan2019lxmert,liSGJXH21,lin2023vision, girdhar2023, zhangLB23} have achieved notable progress in various tasks. However, recent studies \cite{he2020momentum,ma2023robust} found that they also suffer from biases. Our proposed architecture is compatible with these models, thereby improving their robustness.

\section{Dataset Construction and Analysis}
We introduce FortisAVQA, the first dataset designed to assess the robustness of AVQA models. Its construction involves two key processes: rephrasing and splitting. Rephrasing modifies questions from the test set of MUSIC-AVQA to enhance linguistic diversity, thereby mitigating the reliance of models on spurious correlations between key question terms and answers. Splitting entails the automatic and reasonable categorization of questions into frequent (head) and rare (tail) subsets, enabling a more comprehensive evaluation of model performance in both in- and out-of-distribution scenarios.

\subsection{Rephrasing}
\begin{table}[tbp]
\centering
\caption{Statistics of rephrasing consistency. Positive and Negative denote whether the annotator agrees with the rephrasing or not.}
\label{tab:kappa0}
\begin{tabular}{ccc}
\toprule
\textbf{Positive} & \textbf{Negative} & \textbf{Total} \\ \midrule
3                 & 0                 & 164, 219       \\
2                 & 1                 & 47,353         \\
1                 & 2                 & 7,481          \\
0                 & 3                 & 9,172         \\ \bottomrule
\end{tabular}
\end{table}
\begin{figure*}[!ht]
    \centering
    \subfigure{\includegraphics[width=0.6\columnwidth]{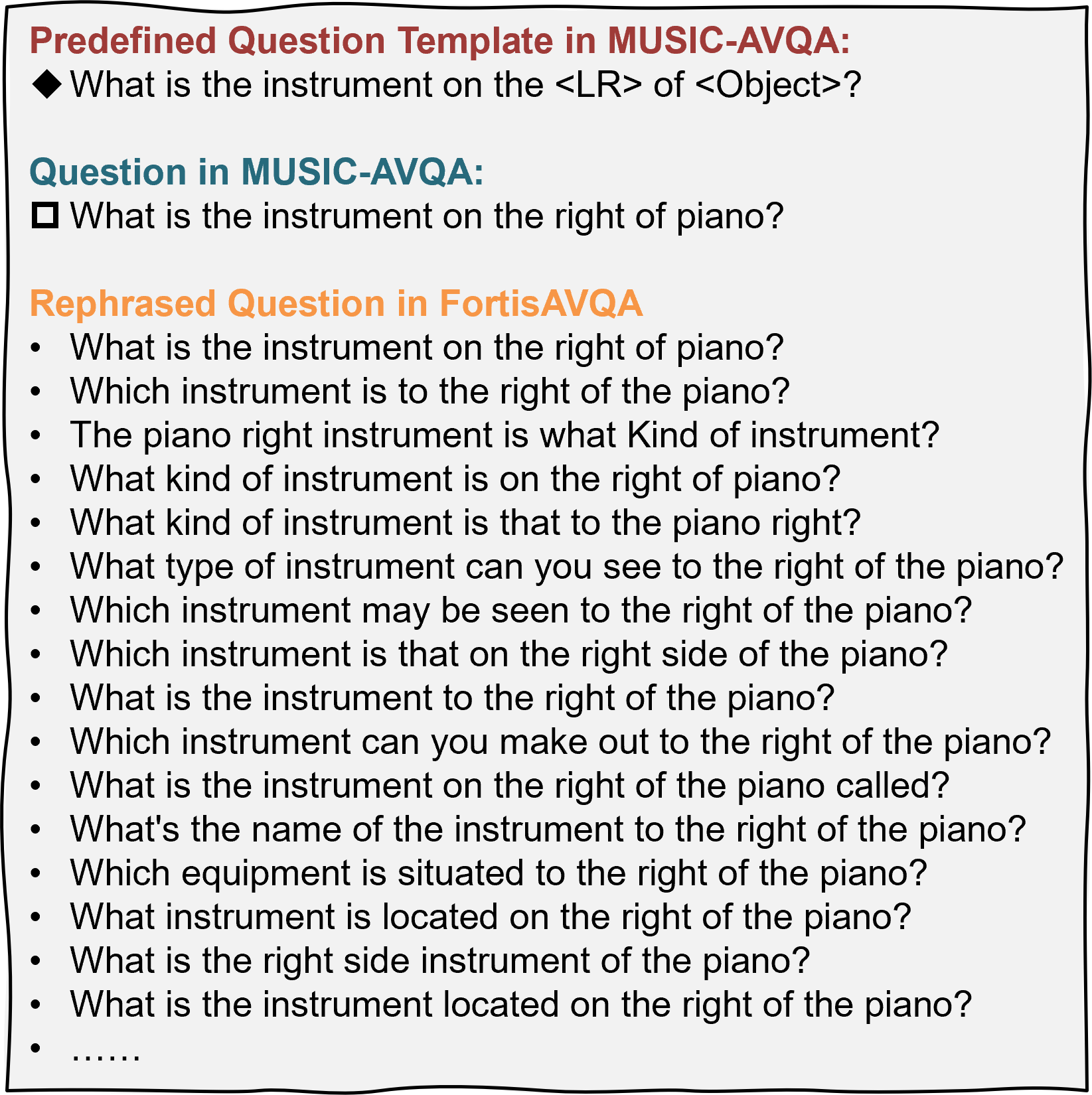}} 
    \hspace{2mm} % 水平间距，根据需要调整
    \subfigure{\includegraphics[width=0.65\columnwidth]{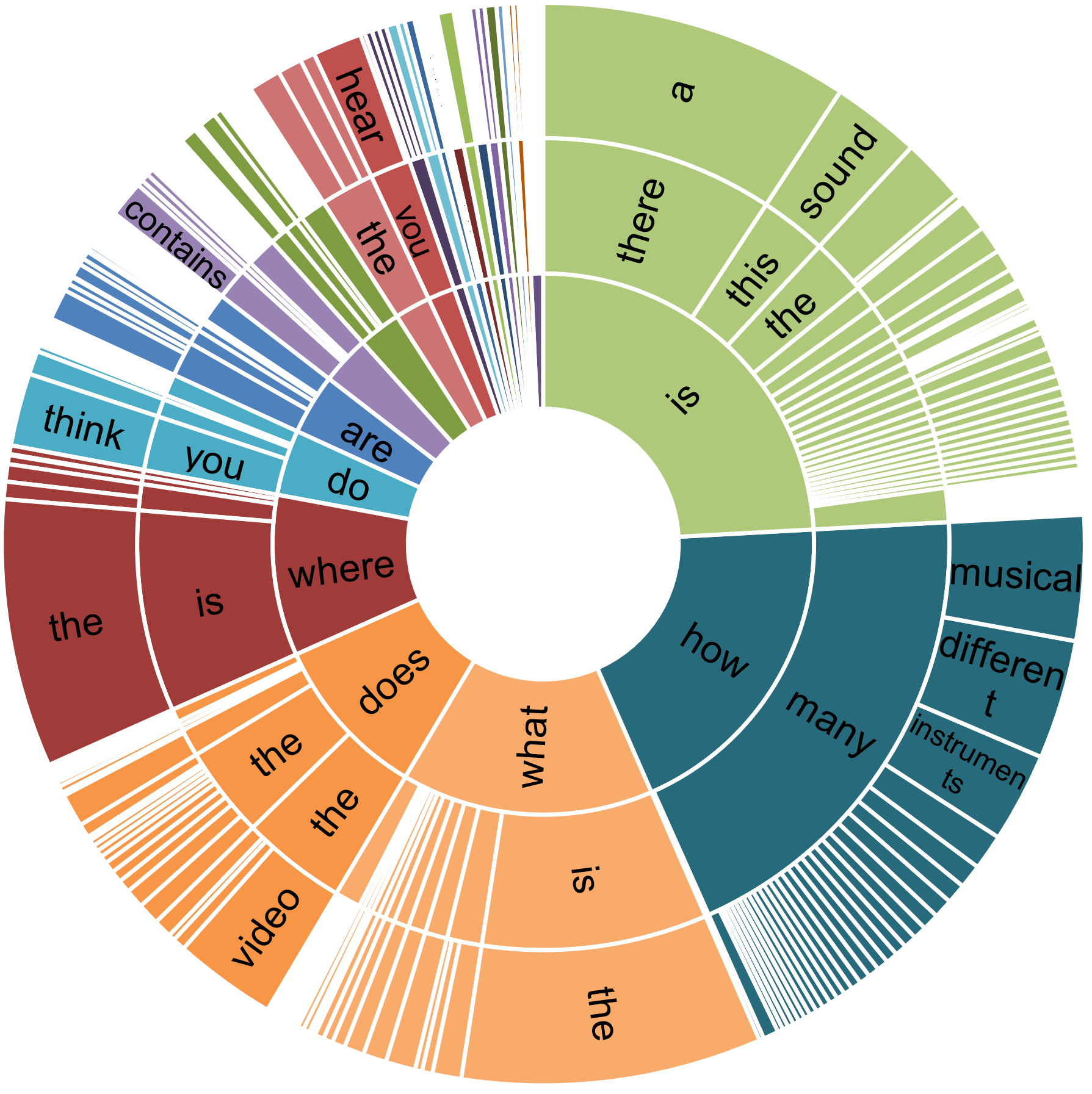}}
    \hspace{2mm} % 水平间距，根据需要调整
    \subfigure{\includegraphics[width=0.65\columnwidth]{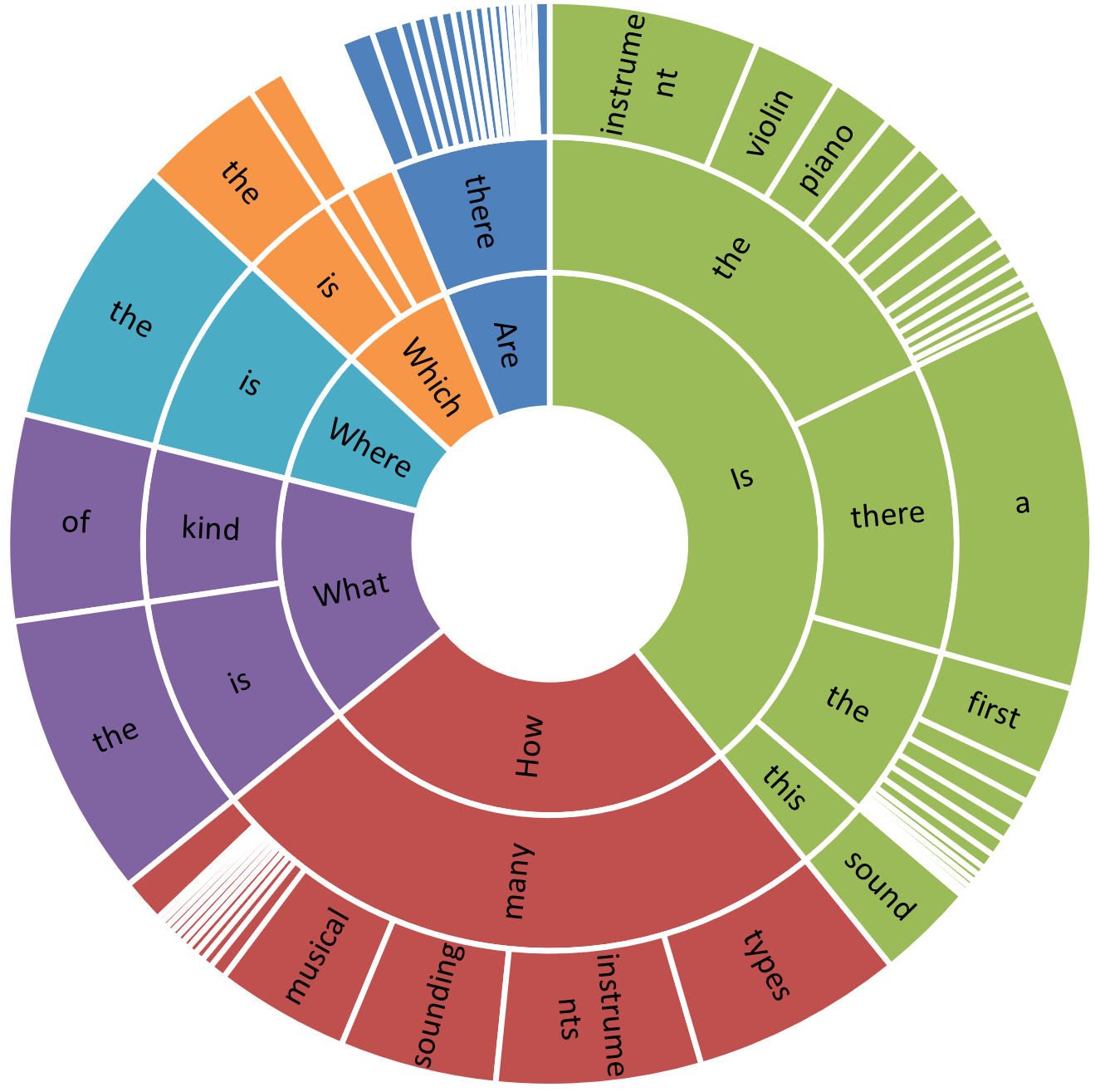}} 
    \caption{Rephrasing visualization of FortisAVQA. The left panel showcases a rephrasing example in FortisAVQA, while the middle and right panels depict the question distribution of FortisAVQA and MUSIC-AVQA, respectively, based on their first three words.}
    \label{fig:dist-v}
\end{figure*}

The questions within the existing dataset \cite{yun2021pano, li2022learning} are formulated using a restricted collection of pre-defined templates. To augment diversity and reality, we employ a rephrasing tool\footnote{\url{https://quillbot.com/paraphrasing-tool}} to rephrase each question 25 times. The left panel of Fig. \ref{fig:dist-v} illustrates an example of rephrasing in FortisAVQA. We observe that the question has more various forms compared to that in MUSIC-AVQA. To ensure the rephrasing quality, three annotators participate in a verification process where their consensus through voting is required. They are all senior students in the field of information science, with one specializing in computer science and the other two in automation. Their extensive professional background equips them with the ability to assess whether the above rephrasing fulfills the requirement. The rephrasing is incorporated into the dataset only when two or more individuals validate the quality of the modifications. According to the statistics, 92.4\% of rephrasing passes this validation, and the Fleiss Kappa value used to measure vote consistency is 0.839. The details are shown in Table \ref{tab:kappa0}. It is evident that the overwhelming majority of the rephrased questions received three favorable votes. These results strongly suggest an exceptionally high quality of the rephrasing efforts.

Fig. \ref{fig:dist-v} presents a comparative analysis of the question distributions in FortisAVQA and MUSIC-AVQA based on the first three words. The results demonstrate the diverse formats of our rephrased questions. Notably, the vocabulary size of our dataset reaches \textbf{465}, which is \textbf{5×} larger than that of MUSIC-AVQA. This substantial increase suggests that our dataset better aligns with real-world linguistic variability. Furthermore, the number of questions in the test split has been significantly expanded from \textbf{9,129} to \textbf{211,572}. This substantial increase in test samples improves the reliability and robustness of evaluations for AVQA methods.

\subsection{Splitting} 
To provide a precise diagnostic for AVQA models, we present a mechanism for introducing distribution shifts based on the answer distribution of specific question types.  This mechanism categorizes the rephrased questions into \emph{head} and \emph{tail}, enabling the assessment of in-distribution and out-of-distribution performance, respectively. We also utilize the \emph{overall} performance to assess model effectiveness on the entire test split.
\begin{table*}[tbp]
\caption{Test split comparison between MUSIC-AVQA and FortisAVQA. EXIST, LOC, CNT, COMP, and TEMP, which are question types, denote `Existential'', Location'', Counting'', Comparative'', and Temporal'', respectively. To mitigate the testing costs of large models, we apply uniform sampling to two datasets at ratios of \(10\%\) and \(1\%\), yielding MUSIC-AVQA\textsubscript{s} and FortisAVQA\textsubscript{s}, respectively. We recommend using the entire test split for evaluating small models, while the sampled split can be used for assessing large multimodal models.}
\label{tab:test-comp}
\centering
\begin{tabular}{c|cc|cc|ccccc|c}
\toprule
\multirow{2}{*}{\textbf{Dataset}} & \multicolumn{2}{c}{\textbf{Audio QA}} & \multicolumn{2}{c|}{\textbf{Visual QA}} & \multicolumn{5}{c|}{\textbf{AVQA}} & \multirow{2}{*}{\textbf{All}}  \\ \cmidrule(lr){2-3} \cmidrule(lr){4-5} \cmidrule(lr){6-10}
  & \multicolumn{1}{c}{\textbf{CNT}} & \textbf{COMP} & \multicolumn{1}{c}{\textbf{CNT}} & \textbf{LOC} & \multicolumn{1}{c}{\textbf{EXIST}} & \multicolumn{1}{c}{\textbf{LOC}} & \multicolumn{1}{c}{\textbf{CNT}} & \multicolumn{1}{c}{\textbf{COMP}} & \textbf{TEMP} & \\ \midrule
MUSIC-AVQA               & \multicolumn{1}{c|}{1,017}     & 594        & \multicolumn{1}{c|}{1,197}     & 1,225     & \multicolumn{1}{c|}{988}         & \multicolumn{1}{c|}{920}      &\multicolumn{1}{c|}{1,265}     & \multicolumn{1}{c|}{1,101}      & 822 & 9,129    \\ 
FortisAVQA       & \multicolumn{1}{c|}{23,107}    & 13,506     & \multicolumn{1}{c|}{27,867}    & 3,3049    & \multicolumn{1}{c|}{25,049}      & \multicolumn{1}{c|}{21,546}   & \multicolumn{1}{c|}{26,565}    & \multicolumn{1}{c|}{23,121}     & 17,762  & 211,572 \\ 
FortisAVQA (Head)    & \multicolumn{1}{c|}{19,033}    & 7,519      & \multicolumn{1}{c|}{20,618}    & 20,949    & \multicolumn{1}{c|}{13,311}      & \multicolumn{1}{c|}{14,890}   & \multicolumn{1}{c|}{22,659}    & \multicolumn{1}{c|}{11,718}     & 11,851  & 142,548 \\ 

FortisAVQA (Tail)    & \multicolumn{1}{c|}{4,074}     & 5,987      & \multicolumn{1}{c|}{7,249}     & 12,100    & \multicolumn{1}{c|}{11,738}      & \multicolumn{1}{c|}{6,656}    & \multicolumn{1}{c|}{3,906}    & \multicolumn{1}{c|}{11,403}     & 5,911  & 69,024  \\ \midrule
MUSIC-AVQA\textsubscript{s} & \multicolumn{1}{c|}{107}    & 61     & \multicolumn{1}{c|}{123}    & 123    & \multicolumn{1}{c|}{100}      & \multicolumn{1}{c|}{93}   & 
                         \multicolumn{1}{c|}{130}    & \multicolumn{1}{c|}{111}     & 85 & 933  \\ 
FortisAVQA\textsubscript{s} & \multicolumn{1}{c|}{233}    & 135     & \multicolumn{1}{c|}{280}    & 331    & \multicolumn{1}{c|}{250}      & \multicolumn{1}{c|}{217}   & 
                         \multicolumn{1}{c|}{268}    & \multicolumn{1}{c|}{231}     & 178 & 2,123  \\ 
FortisAVQA\textsubscript{s} (Head) & \multicolumn{1}{c|}{190}    & 75     & \multicolumn{1}{c|}{206}    & 209    & \multicolumn{1}{c|}{133}      & \multicolumn{1}{c|}{150}   & \multicolumn{1}{c|}{227}    & \multicolumn{1}{c|}{117}     & 118 & 1,425  \\ 
FortisAVQA\textsubscript{s} (Tail) & \multicolumn{1}{c|}{43}    & 60     & \multicolumn{1}{c|}{74}    & 122    & \multicolumn{1}{c|}{117}      & \multicolumn{1}{c|}{67}   & \multicolumn{1}{c|}{41}    & \multicolumn{1}{c|}{114}     & 60  & 698 \\ \bottomrule
\end{tabular}
\end{table*}

\begin{figure*}[tbp]
    \centering
    \includegraphics[width=\textwidth]{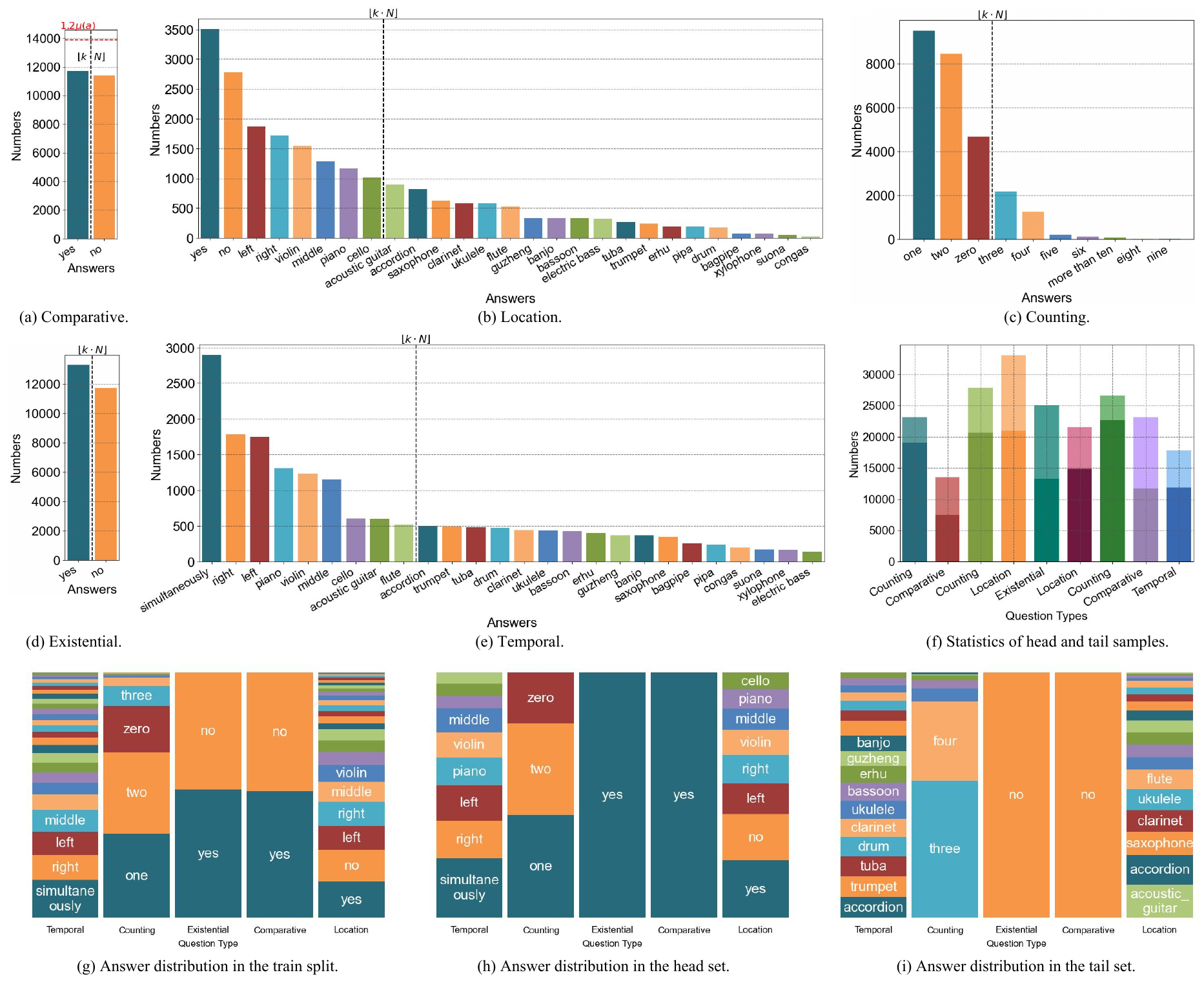}
    \caption{Statistics visualization for the AVQA task in FortisAVQA. $\mu(a)$ is the average number of answers in a group. The previous split mechanism, published in NeurIPS 2024 \cite{ma2024look}, assigns all classes to the tails in subfigure (a). In contrast, the newly proposed approach offers greater flexibility by adapting to the data distribution. $k$ is the ratio in Equation \eqref{constraint}. In the subfigure (f), the dark color denotes the number of head samples, while the light-colored area denotes that of tail samples. In the distribution comparison, we observe a high similarity between train and head sets, whereas the train and tail set exhibit a significant distributional difference. This demonstrates FortisAVQA can evaluate the robustness of multimodal reasoning more comprehensively and precisely compared to existing datasets like MUSIC-AVQA.} \label{fig:fortisavqa}
    \label{fig:dataset}
\end{figure*}

\textbf{Question Grouping} To characterize the distribution shift, we utilize the annotation for question types within MUSIC-AVQA, including ``Existential'', ``Location'', ``Counting'', ``Comparative'', and ``Temporal'', to group questions. Fig. \ref{fig:fortisavqa}(a) to (e) illustrate the answer distribution of type-specific questions within the AVQA task. We see that the answers present a long-tailed distribution. It is essential to note that FortisAVQA encompasses three tasks: audio QA, visual QA, and AVQA. The detailed comparison between MUSIC-AVQA and FortisAVQA is shown in Table \ref{tab:test-comp}. Due to space constraints, we only select AVQA to perform visualization, as shown in Fig. \ref{fig:dataset}.

\textbf{Group Balance Measuring} We characterize the answer balance using Shannon entropy following \cite{kervadec2021roses}, expressed as $H(A) = -\sum_{i=1}^{N}p(a_i) \log p(a_i)$, where $H(A)$ is the entropy of an answer set $A$ for a certain question type, $N$ is the number of answer classes, and $p(a_i)$ represents the probability of answer class $i$. It is important to note that the entropy depends on the number of answer classes, which exhibits significant variability across different question groups. To facilitate meaningful comparisons, we normalize $H(A)$ of each group by $\log (N)$: $\bar{H}(A)=\frac{H(A)}{\log (N)}$, with $\log (N)$ representing the entropy of a uniform distribution of size $N$. Refer to the following proposition for detailed proofs. Thus, the normalized entropy $\bar{H}(A)$ indicates the proximity of the distribution $H(A)$ to a uniform distribution. We preserve the group with normalized entropy below a threshold of 0.9, which aims at selecting imbalanced groups. 

\textit{Proposition.} Given a discrete random variable $X$ that follows a uniform distribution over $N$ possible outcomes, its entropy is defined as $H(X) = \log_2 N$.

\textit{Proof.} 
Let \( X \) be a discrete random variable following a uniform distribution over a finite support of size \( N \). The entropy \( H(X) \) is defined as:
\begin{align}
    H(X) &= -\sum_{i=1}^{N} p(x_i) \log_2 p(x_i),
\end{align}
where \( p(x_i) \) denotes the probability mass function evaluated at \( x_i \).

Since \( X \) follows a uniform distribution, each outcome occurs with equal probability: 
\begin{align}
    p(x_i) = f(x_i) = \frac{1}{N}, \quad \forall i \in \{1,2,\dots,N\}.
\end{align}

Substituting this into the entropy definition, we obtain:
\begin{align}
    H(X) &= -\sum_{i=1}^{N} \frac{1}{N} \log_2 \left( \frac{1}{N} \right), \\
    &= -\frac{1}{N} \log_2 \left( \frac{1}{N} \right) \sum_{i=1}^{N} 1, \\
    &= -\frac{1}{N} \log_2 \left( \frac{1}{N} \right) \cdot N, \\
    &= -\log_2 \left( \frac{1}{N} \right), \\
    &= \log_2 N.
\end{align}

\textbf{In- and Out-of-Distribution Splitting} In our previous work \cite{ma2024look}, we define the \emph{tail} class as class $i$ with $|a_i| \leq 1.2 \mu(a)$, where $|a_i|$ represents the number of samples belonging to answer class $i$, and $\mu(a)$ denotes the average sample count for a group. While the empirical study provides an intuitive threshold, it introduces a critical limitation: The fixed multiplier (i.e., $1.2\times$) lacks adaptability to varying data distributions. For example, Fig. \ref{fig:fortisavqa}(a) shows that this criterion erroneously classifies all classes as \emph{tail} due to their similar sample counts.

To address the mentioned issue, we employ a conformal-inspired optimization \cite{oliveira2024split} that dynamically adjusts the \emph{head and tail} boundary. An ideal split should satisfy two competing objectives, i.e., \textit{Coverage} and \textit{Compactness}. Coverage indicates that \emph{head} classes must contain sufficient samples and Compactness suggests that the \emph{head} set should be as compact as possible. This duality aligns perfectly with the core philosophy of conformal prediction, enabling distribution-free control over the split. %which balances coverage validity with the minimization of the size of the prediction set.}

Technically, we formulate \emph{head} selection as constrained optimization, which controls the trade-off between \emph{head} class proportion and coverage guarantee. Formally, given a dataset $\mathcal{D}$  with $N$ total classes, we determine the optimal \emph{head} class $\mathcal{H} \subset \{1,...,N\} $ by solving:
\begin{align}
\label{optimize}
& \underset{k}{\text{minimize}} & & |\mathcal{H}| = \lfloor kN \rfloor, \\
\label{constraint}
& \text{subject to} & & \frac{\sum_{c\in\mathcal{H}}n_c}{|\mathcal{D}|} \geq 1 - k,
\end{align}
where $n_c$ denotes the sample count of class $c$. This formulation directly mirrors conformal prediction on:
\begin{itemize}
    \item \textbf{Coverage Guarantee.} The constraint in Equation \eqref{constraint} enforces that head classes must collectively contain at least $1-k$ proportion of the dataset, which corresponds to the confidence requirement in conformal prediction.
    \item \textbf{Compactness.} Minimizing $k$ reduces $|\mathcal{H}|$, which aligns with conformal prediction's goal of minimizing prediction set sizes under coverage guarantees.
\end{itemize}
\begin{figure*}[tbp]
    \centering  %图片全局居中
    \includegraphics[width=0.9\textwidth]{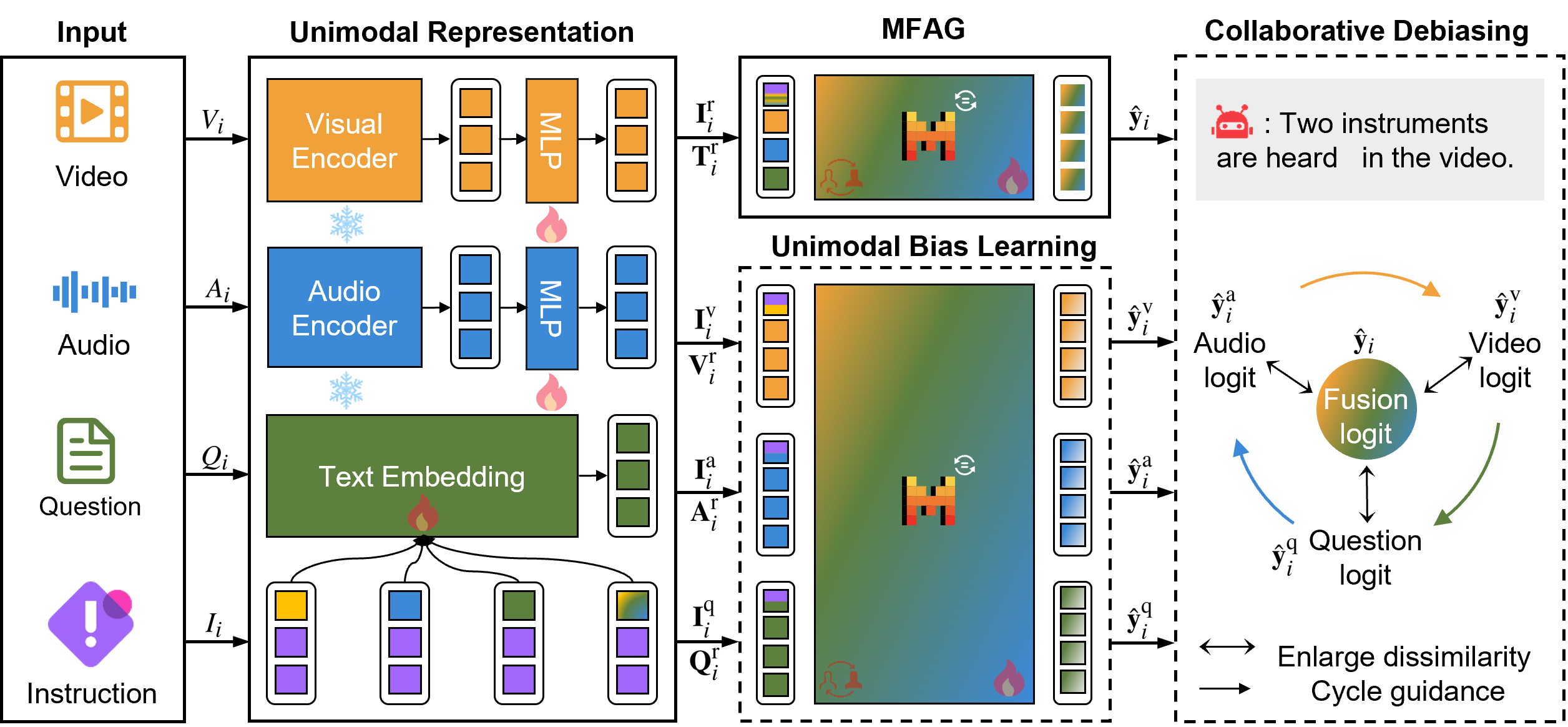}
    \caption{Illustration of our proposed Multimodal Audio-Visual Epistemic Network (MAVEN). The instructions can be classified into two categories: multimodal fusion and modality-specific bias learning. MFAG denotes the multimodal fusion and answer generation. During the test stage, the module marked with dash lines is removed. \includegraphics[height=0.8\baselineskip]{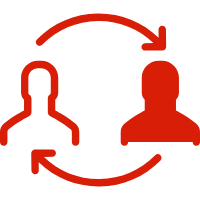} denotes the parameter sharing. \includegraphics[height=0.8\baselineskip]{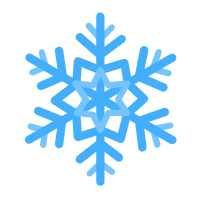} represents the parameter is frozen. \includegraphics[height=0.8\baselineskip]{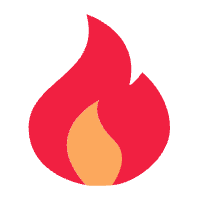} indicates that the parameter is trainable. }
    \label{fig:model}
\end{figure*}

The dynamic optimization automatically adjusts the boundary through the $k$-dependent constraint, thereby eliminating the dependency on data distribution. Therefore, the proposed split mechanism ensures precise controls over the proportion of samples captured by \emph{head} classes. The aforementioned rephrasing and splitting procedures are only carried out in the test split of MUSIC-AVQA. Fig. \ref{fig:fortisavqa}(g) to (i) illustrate the answer distribution comparison between the training and test splits. We observe that the distribution of the training and head splits is relatively similar, whereas the distribution between the training and test splits differs significantly. This experimental setup effectively captures model performance in both in-distribution and out-of-distribution scenarios. Consequently, our proposed dataset, FortisAVQA, facilitates a more precise and comprehensive evaluation of models in addressing data biases.

%\begin{figure}[tbp]
%    \centering  %图片全局居中
%    \includegraphics[width=0.8\columnwidth]{images/dataset/head_tail_new.pdf}
%    \caption{Statistics of head and tail samples under specific types of questions in each task. The dark color denotes the number of head samples, while the light-colored area denotes that of tail samples.}
%    \label{fig:head-tail}
%\end{figure}

\section{Methodlogy}
\label{sec:method}
Existing methods \cite{li2022learning,lao2023coca,fu2024vita} suffer from harmful bias in the training split, resulting in high in-distribution but low out-of-distribution performance. To address this issue, we propose a robust generative framework, MAVEN, which employs a multifaceted cycle collaborative debiasing strategy. As illustrated in Fig. \ref{fig:model}, MAVEN first extracts unimodal embeddings using modality-specific encoders. It then fine-tunes a parameter-sharing generative model to learn multimodal fusion while capturing three unimodal biases, guided by four distinct prompts. Finally, a collaborative debiasing strategy is applied to amplify the disparity between the fusion logit and the bias logit. Simultaneously, a cycle-guidance mechanism is introduced to maintain consistency among bias logit.

\textbf{Unimodal Representation.} Given an AVQA sample consisting of a video and its corresponding question, we begin by segmenting the video into non-overlapping visual and audio pairs, each spanning one second. The visual and audio sequences are denoted as $V_i$ and $A_i$, respectively. Subsequently, a distinct embedding module is employed to acquire unimodal embeddings. Specifically, we utilize the fixed InternViT-300M-448px\footnote{\url{https://huggingface.co/OpenGVLab/InternViT-300M-448px}} model as the visual encoder. This model supports a dynamic input resolution of $448 \times 448$ with a basic tile size of $448 \times 448$, generating 256 token embeddings. For audio processing, we employ a Mel Filter Bank block, which decomposes the audio segment into individual frequency bands based on the Mel scale—a perceptual scale designed to reflect the nonlinear characteristics of human auditory perception. The extracted audio features are then passed through four convolutional neural network downsampling layers, which effectively reduce the input dimensionality while preserving crucial temporal and spectral information. Subsequently, the processed audio features are fed into a stack of 24 Transformer layers with fixed parameters. This module is pre-trained on large-scale speech recognition and audio captioning datasets \cite{fu2024vita} and is capable of capturing long-range dependencies in the audio sequence while modeling complex temporal relationships. To bridge the video/audio and text modalities, we introduce two trainable Multi-Layer Perceptrons (MLPs) as modality adapters, mapping the video and audio features into $d_r$-dimensional spaces. Finally, we obtain the video and audio representations $\mathbf{V}_i^\mathrm{r} \in \mathbb{R}^{T_v \times d_r}$ and $\mathbf{A}_i^\mathrm{r} \in \mathbb{R}^{T_a \times d_r}$, where $T_v$ represents the number of tokens extracted from the sampled video frames, and $T_a$ denotes the number of tokens generated from the audio signal.

% Finally, each 2-second segment of audio input is encoded into 25 tokens, which serve as the compact representation $\mathbf{A}_i^\mathrm{r} \in \mathbb{R}^{T_a \times d_r}$ of the audio that can be further utilized in the following module. Here, $T_a$ is the number of tokens generated by the audio segments.

\textbf{Multimodal Fusion and Answer Generation.} Previous works \cite{ma2024look,lin2023vision,li2022learning} have predominantly treated AVQA as a discrimination task, which may limit its practical applicability. In real-world scenarios, AVQA typically requires models to simultaneously comprehend video, audio, and query inputs before generating answers. To the best of our knowledge, no publicly available large models support all three modalities as input simultaneously. Instead, existing models \cite{cheng2024videollama,lu2024unified} generally process only two modalities at a time. For instance, VITA \cite{fu2024vita} is a multimodal large language model designed to handle video, image, text, and audio inputs. However, it only supports the simultaneous processing of query inputs (either text or audio) and visual data. To overcome this limitation, we first treat the features of the three modalities as token embeddings $\mathbf{E}_i^{\mathrm{r}} \in \mathbb{R}^{(T_a+T_v+T_q) \times d_r}$, incorporating special tokens at the beginning of each modality, where $T_q$ is the number of tokens of questions. Then, we perform instruction tuning for VITA to enhance its capability for simultaneously integrating audio, video, and text. After multimodal fusion $g$, the generated answer logit $\hat{\mathbf{y}}_i \in \mathbb{R}^{T_g \times d_a}$ is obtained, where $T_g$ represents the maximum number of generated tokens, and $d_a$ denotes the vocabulary size. The answer-generation process is formulated as follows:
\begin{align}
    p(\hat{\mathbf{y}}_i \mid \mathbf{I}_i^\mathrm{r}, \mathbf{Q}_i^\mathrm{r}, \mathbf{V}_i^\mathrm{r}, \mathbf{A}_i^\mathrm{r}) &= \prod_{t=1}^{l_a} p(\hat{\mathbf{y}}_{i,t} \mid \hat{\mathbf{y}}_{i,<t}, \mathbf{M}_i), \\
    \mathbf{M}_i &= g(\mathbf{I}_i^\mathrm{r}, \mathbf{E}_i^{\mathrm{r}}), \\
    \mathbf{E}_i^{\mathrm{r}} &= [\mathbf{Q}_i^\mathrm{r} || \mathbf{V}_i^\mathrm{r}|| \mathbf{A}_i^\mathrm{r}],
\end{align}
where $\mathbf{I}_i^\mathrm{r} \in \mathbb{R}^{T_f \times d_r}$ represents the token-level embedding of multimodal fusion and answer generation instructions with the length $T_f$, $\mathbf{M}_i \in \mathbb{R}^{(T_f+T_a+T_v+T_q) \times d_r}$ denotes the multimodal representation, $\hat{\mathbf{y}}_{i,<t} = (\hat{\mathbf{y}}_{i,1}, \hat{\mathbf{y}}_{i,2}, \dots, \hat{\mathbf{y}}_{i,t-1})$ is the previously generated sequence representation, and $||$ denotes the concatenation operation.

\textbf{Unimodal Bias Learning.} AVQA may suffer from harmful unimodal biases, encompassing biases associated with audio, video, and language (question), respectively. To capture these unimodal biases, we employ a bias learner that processes only one of the three modalities at a time while sharing parameters with the aforementioned answer generation module. The answer logit $\hat{\mathbf{y}}_i^{\mathrm{v}} \in \mathbb{R}^{T_g \times d_a}$, which depends solely on the video input, is computed as follows:
\begin{align}
    p(\hat{\mathbf{y}}_i^{\mathrm{v}} \mid \mathbf{I}_i^\mathrm{v}, \mathbf{V}_i^\mathrm{r}) &= \prod_{t=1}^{l_a} p(\hat{\mathbf{y}}_{i,t}^{\mathrm{v}} \mid \hat{\mathbf{y}}_{i,<t}^{\mathrm{v}}, \mathbf{M}_i^\mathrm{v}), \\
    \mathbf{M}_i^{\mathrm{v}} &= g(\mathbf{I}_i^\mathrm{v}, \mathbf{V}_i^\mathrm{r}),
\end{align}
where $\mathbf{I}_i^\mathrm{v}$ is the embedding of the instruction that prompts MAVEN to generate answers only depending on the video input. Similarly, the answer logit $\hat{\mathbf{y}}_i^{\mathrm{a}}, \hat{\mathbf{y}}_i^{\mathrm{q}} \in \mathbb{R}^{T_g \times d_a}$, which depend only on audio and language inputs, respectively, are obtained in the same manner. Notably, these bias learners are removed during the testing phase. The instructions for multimodal fusion and bias learning are shown in Fig. \ref{fig:case}.

\textbf{Collaborative Debiasing.} To eliminate bias learning, we propose a \textit{multifaceted cycle collaborative debiasing} (MCCD) strategy. It first reduces the bias impact from multiple views by enlarging the dissimilarity between unimodal and multimodal logit. This discrepancy enlargement $\mathcal{L}_{\mathrm{d}}$ is implemented by the KL divergence:
\begin{align} 
\mathcal{L}_{\mathrm{d}} &= \alpha \sum_{k \in \{\mathrm{q}, \mathrm{v}, \mathrm{a} \} } \frac{1}{D_{\mathrm{KL}}(\hat{\mathbf{y}}_i \parallel \hat{\mathbf{y}}_i^k)}, \label{eq:dist} \\ 
D_{\mathrm{KL}}(\hat{\mathbf{y}}_i \parallel \hat{\mathbf{y}}_i^k) &= \sum_i \hat{\mathbf{y}}_i \log \frac{\hat{\mathbf{y}}_i}{\hat{\mathbf{y}}_i^{k}}, 
\end{align}
where $\alpha$ denotes the factor to adjust weight, $D_{\mathrm{KL}}(\hat{\mathbf{y}}_i \parallel \hat{\mathbf{y}}_i^k)$ measures the distribution difference between unimodal and multimodal logit and $\epsilon=1e{-5}$ is added to the denominator to avoid division by zero. Please note that the input of $D_{\mathrm{KL}}$ is activated by a $\mathrm{softmax}$ function.

There are multiple possible directions to enhance the differentiation among high-dimensional vectors in the above equation. For instance, increasing the disparity in the fourth dimension between the visual and multimodal logits can amplify their overall difference. To ensure a stable optimization, a constraint may be added to minimize $L_d$. Intuitively, relying solely on one modality for answer prediction may result in similar unimodal logit distributions. Therefore, MAVEN employs cycle guidance to regulate the logit distribution. This guidance $\mathcal{L}_{\mathrm{c}}$ is also implemented by the KL divergence:
\begin{align}
    \mathcal{L}_\mathrm{c} = \beta \sum_{(j, k) \in \{\mathrm{(l,a), (a,v), (v,l)}\}} D_{\mathrm{KL}}(\hat{\mathbf{y}}_i^{j} \parallel \hat{\mathbf{y}}_i^{k}), \label{eq:cyc}
\end{align}
where $\beta$ is the factor to control weight, $D_{\mathrm{KL}}(\hat{\mathbf{y}}_i^{j} \parallel \hat{\mathbf{y}}_i^{k})$ denotes the relative entropy between unimodal logit.

Finally, the total objective function is formulated as the summation of \(\mathcal{L}_{\mathrm{d}}\), \(\mathcal{L}_{\mathrm{c}}\), and \(\mathcal{L}_{\mathrm{a}}\), which collectively optimize the MAVEN parameters. \(\mathcal{L}_{\mathrm{a}}\) represents the answer generation loss, defined as follows:  
\begin{align}
    \mathcal{L}_{\mathrm{a}} = - \sum_i \mathbf{y}_i \log \hat{\mathbf{y}}_i,
\end{align}
where \(\mathbf{y}_i\) and \(\hat{\mathbf{y}}_i\) denote the one-hot token-level ground-truth answer label and the predicted logit obtained from multimodal fusion, respectively.
The training procedure of MAVEN is outlined in Algorithm \ref{ag:mt}, where $L$ represents the number of training samples. The audio encoder $f_a$ has fixed parameters, while $f_m$ and the $\mathrm{BiasLearner}$ share the same set of parameters. During the testing phase, the bias learner is removed. 
\begin{algorithm}[t]
\caption{Model Training} \label{ag:mt}
\KwIn{$\mathcal{D}=\{(A_i, V_i, Q_i, y_i)\}_{i=1}^{L}$.}
\KwOut{MAVEN}
Initialize model parameters $\theta$\;
Initialize Adam optimizer\;
Set learning rate $\eta$\;
Set the number of training epochs $M$\;
\For{$epoch \gets 1$ to $M$}{
  \For{each batch in $\mathcal{D}$}{
    \textbf{Learn unimodal representations: }
    $\mathbf{V}_i^{\mathrm{r}} \gets \mathrm{MLP}_{\mathrm{v}}(\mathrm{InternViT}(V_i))$
    $\mathbf{A}_i^{\mathrm{r}} \gets \mathrm{MLP}_\mathrm{a}(f_a(A_i))$ \\
    \textbf{Multimodal Fusion and Answer generation:}
    $\hat{\mathbf{y}}_i \gets f_{\mathrm{m}}(\mathbf{I}_i^\mathrm{r}, \mathbf{Q}_i^\mathrm{r}, \mathbf{V}_i^\mathrm{r}, \mathbf{A}_i^\mathrm{r})$ \\
    \textbf{Capture unimodal biases: }
    $\hat{\mathbf{y}}_i^{\mathrm{a}} \gets \mathrm{BiasLearner}_{\mathrm{a}}(\mathbf{I}_i^\mathrm{a}, \mathbf{A}_i^{\mathrm{r}})$
    $\hat{\mathbf{y}}_i^{\mathrm{v}} \gets \mathrm{BiasLearner}_{\mathrm{v}}(\mathbf{I}_i^\mathrm{v}, \mathbf{V}_i^{\mathrm{r}})$
    $\hat{\mathbf{y}}_i^{\mathrm{q}} \gets \mathrm{BiasLearner}_{\mathrm{q}}(\mathbf{I}_i^\mathrm{q}, \mathbf{Q}_i^{\mathrm{r}})$ \\
    \textbf{Joint Loss Computation: } 
    $\mathcal{L}_{\mathrm{a}} \gets - \sum \mathbf{y}_i \log \hat{\mathbf{y}}_i$
    $\mathcal{L}_{\mathrm{d}}, \mathcal{L}_{\mathrm{c}} \gets \mathrm{MCCD}(\hat{\mathbf{y}}_i^{\mathrm{a}}, \hat{\mathbf{y}}_i^{\mathrm{v}}, \hat{\mathbf{y}}_i^{\mathrm{q}},\hat{\mathbf{y}}_i^{\mathrm{m}})$
    $\mathcal{L} \gets \mathcal{L}_{\mathrm{a}} + \mathcal{L}_{\mathrm{d}} + \mathcal{L}_{\mathrm{c}}$ \\
    \textbf{Parameter Update: } \\
    ~~$\nabla \theta \gets \nabla_{\theta} \mathcal{L}$ \\
    ~~$\theta \gets \theta - \eta \cdot \nabla \theta$
  }
}
\KwRet{$\mathrm{MAVEN}$}
\end{algorithm}

\section{Experiments}
%We first conduct experiments on the MUSIC-AVQA dataset to verify the effectiveness of our proposed architecture. Subsequently, we leverage the proposed benchmark to reevaluate existing AVQA methods. Following that, we conduct ablation studies to explore the debiasing contribution of each component within the MCCD strategy. Finally, we present visualizations and conduct an in-depth analysis of the qualitative results.

\subsection{Dataset and Evaluation}
\begin{table*}[!ht]
\centering
\caption{Experimental results (\%) on the MUSIC-AVQA test split. F represents whether the MCCD strategy is integrated. EXIST, LOC, CNT, COMP, and TEMP, which are question types, denote ``Existential'', ``Location'', ``Counting'', ``Comparative'', and ``Temporal'', respectively. Avg. denotes the average accuracy.}
\label{tab:exp-music}

\begin{tabular}{ccccccccccccccc}
\toprule
\multirow{2}{*}{\textbf{Method}} & \multirow{2}{*}{\textbf{F}} & \multicolumn{3}{c}{\textbf{Audio QA}} & \multicolumn{3}{c}{\textbf{Visual QA}} & \multicolumn{6}{c}{\textbf{AVQA}}  & \textbf{All}   \\ \cmidrule(lr){3-5} \cmidrule(lr){6-8} \cmidrule(lr){9-14} \cmidrule(lr){15-15}
    &    & \textbf{CNT} & \textbf{COMP} & \textbf{Avg.} & \textbf{CNT} & \textbf{LOC} & \textbf{Avg.} & \textbf{EXIST} & \textbf{LOC} & \textbf{CNT} & \textbf{COMP} & \textbf{TEMP} & \textbf{Avg.} & \textbf{Avg.}  \\ \midrule
\multirow{2}{*}{FCNLSTM}  & \ding{55}  & 69.96  & 61.06  & 66.67  & 63.89  & 58.14  & 60.98  & 83.42   & 56.31  & 60.28  & 50.85  & 56.92  & 61.46  & 62.25   \\
                          & \ding{51}  & 69.57  & 61.90  & 66.73  & 63.30  & 58.47  & 60.86  & 83.82   & 56.74  & 60.13  & 51.57  & 56.92  & 61.73  & \textbf{62.38} \\
\multirow{2}{*}{CONVLSTM} & \ding{55}  & 68.88  & 63.06  & 66.73  & 64.89  & 58.55  & 61.68  & 82.81   & 55.99  & 61.30  & 53.45  & 54.73  & 61.75  & 62.61   \\
                          & \ding{51}  & 70.06  & 64.56  & 68.02  & 65.89  & 58.79  & 62.30  & 82.81   & 57.17  & 61.77  & 54.62  & 57.16  & 62.73  & \textbf{63.55} \\ \midrule
\multirow{2}{*}{MCAN}     & \ding{55}  & 75.05  & 54.58  & 67.47  & 68.06  & 72.15  & 70.13  & 81.91   & 54.15  & 53.45  & 52.11  & 47.21  & 57.80  & 62.77   \\
                          & \ding{51}  & 76.52  & 55.74  & 68.82  & 69.64  & 72.15  & 70.91  & 80.70   & 53.94  & 64.91  & 53.54  & 59.71  & 62.69  & \textbf{65.94} \\
\multirow{2}{*}{GRU}      & \ding{55}  & 71.82  & 58.90  & 67.04  & 66.06  & 71.82  & 68.97  & 81.41   & 60.30  & 62.32  & 56.23  & 61.89  & 64.26  & 66.00   \\ 
                          & \ding{51}  & 71.82  & 59.23  & 67.16  & 67.81  & 71.42  & 69.63  & 82.31   & 60.73  & 63.81  & 55.16  & 62.26  & 64.71  & \textbf{66.45} \\ \midrule
\multirow{2}{*}{HCRN}     & \ding{55}  & 70.21  & 45.62  & 61.14  & 62.41  & 51.51  & 56.90  & 52.94   & 42.07  & 54.70  & 50.59  & 33.33  & 48.41  & 52.54   \\
                          & \ding{51}  & 73.16  & 58.25  & 67.66  & 69.84  & 60.98  & 65.36  & 80.87   & 51.09  & 63.64  & 55.11  & 60.94  & 60.24  & \textbf{64.31} \\ 
\multirow{2}{*}{PSAC}     & \ding{55}  & 71.33  & 56.07  & 65.68  & 65.89  & 72.07  & 69.02  & 78.59   & 54.80  & 63.11  & 55.96  & 61.17  & 62.75  & 64.92   \\
                          & \ding{51}  & 72.02  & 60.40  & 67.71  & 65.39  & 71.34  & 68.40  & 80.30   & 54.05  & 63.19  & 56.50  & 60.80  & 63.02  & \textbf{65.27} \\ \midrule
\multirow{2}{*}{LAViT}    & \ding{55}  & 74.27  & 65.06  & 70.86  & 69.89  & 77.12  & 73.55  & 81.21   & 62.03  & 65.93  & 60.90  & 63.96  & 66.78  & 69.29   \\
                          & \ding{51}  & 74.36  & 65.06  & 70.92  & 68.72  & 77.04  & 72.93  & 81.61   & 60.73  & 66.01  & 63.68  & 63.59  & 67.19  & \textbf{69.36} \\
\multirow{2}{*}{LAVisH}   & \ding{55}  & 78.18  & 58.74  & 70.98  & 75.65  & 78.75  & 77.21  & 81.41   & 63.54  & 71.98  & 57.76  & 66.38  & 68.30  & 71.13   \\
                          & \ding{51}  & 76.13  & 58.57  & 69.62  & 74.81  & 77.28  & 76.06  & 81.71   & 64.19  & 70.64  & 59.91  & 66.26  & 68.59  & 70.75   \\ \midrule
\multicolumn{1}{l}{VideoLLaMA 2} & \ding{55}  & 79.44  & 52.46  & 69.64  & 81.30  & 82.93  & 82.11  & 77.00   & 63.44  & 77.69  & 59.46  & 64.71  & 68.98  & 72.56   \\
\multirow{2}{*}{MAVEN}    & \ding{55}  & 79.44  & 54.10  & 72.79  & 80.49  & 93.50  & 86.99  & 87.00   & 66.67  & 73.85  & 54.95  & 68.24  & 69.94  & 74.60   \\
                          & \ding{51}  & 78.50  & 55.74  & 70.24  & 82.93  & 93.50  & 88.21  & 86.00   & 64.52  & 76.92  & 64.86  & 68.24  & 72.45  & \textbf{76.21} \\ 
VITA                      & \ding{55}  & 59.81  & 45.90  & 54.76  & 50.41  & 34.96  & 42.68  & 54.00   & 49.46  & 46.92  & 27.93  & 41.18  & 43.74  & 45.44   \\
Qwen2.5-VL                & \ding{55}  & 48.60  & 55.00  & 51.80  & 55.28  & 53.66  & 54.47  & 44.00   & 52.17  & 63.57  & 37.84  & 41.18  & 47.75  & 50.14   \\
GPT 4o                    & \ding{55}  & 65.42  & 36.07  & 50.75  & 72.36  & 62.30  & 67.33  & 56.12   & 54.84  & 59.23  & 37.84  & 42.35  & 50.08  & 54.06  \\ \bottomrule
\end{tabular}

\end{table*}
\begin{table*}[t]
\caption{Experimental results (\%) on the FortisAVQA test split. The question types, such as CNT and COMP, are introduced in Table \ref{tab:exp-music}. H and T denote the head and tail accuracy. There is no publicly available code for COCA.}
\label{tab:music-r}
\centering
\resizebox{\textwidth}{!}{
\begin{tabular}{ccccccccccccccccccccc}
\toprule
\multirow{3}{*}{Method}   & \multirow{3}{*}{F} & \multicolumn{4}{c}{Audio QA} & \multicolumn{4}{c}{Visual QA} & \multicolumn{10}{c}{AVQA} & All \\ \cmidrule(lr){3-6} \cmidrule(lr){7-10} \cmidrule(lr){11-20} \cmidrule(lr){21-21}
&  & \multicolumn{2}{c}{CNT} & \multicolumn{2}{c}{COMP} & \multicolumn{2}{c}{CNT} & \multicolumn{2}{c}{LOC} & \multicolumn{2}{c}{EXIST} & \multicolumn{2}{c}{LOC} & \multicolumn{2}{c}{CNT} & COMP  &   & \multicolumn{2}{c}{TEMP} & \multirow{2}{*}{Avg.} \\ \cmidrule(lr){3-4} \cmidrule(lr){5-6} \cmidrule(lr){7-8} \cmidrule(lr){9-10} \cmidrule(lr){11-12} \cmidrule(lr){13-14} \cmidrule(lr){15-16} \cmidrule(lr){17-18} \cmidrule(lr){19-20}
&   & H & T & H  & T & H & T & H & T & H  & T  & H & T & H & T & H     & T     & H  & T     &  \\ \midrule
\multirow{2}{*}{FCNLSTM}  & \ding{55} & 66.23 & 36.48 & 64.78  & 51.14 & 62.61 & 34.96 & 55.32 & 48.79 & 64.76  & 78.52  & 49.93 & 45.87 & 51.26 & 7.04  & 43.13 & 71.67 & 37.55  & 27.90 & 54.11  \\
& \ding{51} & 63.31 & 39.15 & 58.36  & 59.98 & 63.76 & 32.25 & 55.30 & 47.68 & 67.56  & 74.60  & 51.58 & 46.29 & 50.86 & 6.68  & 44.64 & 65.82 & 40.22  & 31.04 & 53.92  \\
\multirow{2}{*}{CONVLSTM} & \ding{55} & 70.47 & 40.77 & 67.80  & 53.15 & 62.77 & 37.29 & 54.15 & 47.83 & 59.99  & 83.63  & 50.70 & 40.93 & 48.99 & 8.09  & 43.25 & 71.50 & 42.88  & 42.67 & 55.04  \\ 
 & \ding{51} & 67.32 & 56.47 & 56.47  & 62.90 & 62.15 & 33.94 & 60.04 & 49.88 & 64.66  & 79.72  & 51.23 & 43.96 & 48.25 & 7.68  & 54.31 & 61.61 & 42.25  & 47.67 & \textbf{55.76}   \\ \midrule
\multirow{2}{*}{MCAN}& \ding{55} & 70.82 & 59.28 & 50.71  & 56.14 & 61.72 & 47.99 & 65.91 & 64.62 & 58.17  & 60.67  & 51.13 & 51.22 & 44.31 & 38.91 & 66.00 & 36.80 & 36.70  & 58.77 & 55.80  \\
 & \ding{51} & 75.26 & 60.46 & 47.99  & 59.31 & 67.86 & 59.78 & 65.63 & 45.34 & 60.50  & 59.15  & 48.93 & 50.15 & 51.28 & 39.30 & 53.90 & 58.19 & 40.00  & 62.82 & \textbf{57.49}   \\
\multirow{2}{*}{GRU} & \ding{55} & 68.08 & 49.12 & 63.60  & 55.95 & 69.74 & 27.34 & 62.01 & 60.17 & 80.29  & 56.88  & 43.21 & 28.19 & 52.31 & 14.90 & 56.39 & 60.50 & 32.90  & 35.29 & 55.37  \\
 & \ding{51} & 63.75 & 48.87 & 69.54  & 48.24 & 69.60 & 31.94 & 65.99 & 60.20 & 82.00  & 51.98  & 44.10 & 39.38 & 52.00 & 13.13 & 53.16 & 60.96 & 36.49  & 38.96 & \textbf{55.84}   \\ \midrule
\multirow{2}{*}{HCRN}& \ding{55} & 54.01 & 50.83 & 35.08  & 42.19 & 45.62 & 22.32 & 37.08 & 51.34 & 41.48  & 66.95  & 35.49 & 28.98 & 46.12 & 26.01 & 39.00 & 45.34 & 32.13  & 36.97 & 42.66  \\
 & \ding{51} & 58.62 & 43.40 & 22.54  & 50.54 & 48.37 & 20.73 & 38.03 & 49.74 & 44.93  & 59.82  & 36.62 & 30.15 & 46.53 & 24.65 & 40.56 & 28.57 & 42.43  & 43.04 & \textbf{42.81}   \\
\multirow{2}{*}{PSAC}& \ding{55} & 56.84 & 48.67 & 61.83  & 45.11 & 54.75 & 31.58 & 69.23 & 65.38 & 56.43  & 58.53  & 42.67 & 45.94 & 45.16 & 31.08 & 40.08 & 62.36 & 29.79  & 49.99 & 51.80  \\
 & \ding{51} & 58.41 & 46.34 & 58.62  & 55.44 & 52.83 & 34.58 & 69.74 & 65.28 & 56.31  & 54.61  & 43.18 & 43.72 & 43.42 & 29.13 & 46.35 & 58.79 & 27.66  & 49.31 & 51.57  \\ \midrule
\multirow{2}{*}{LAViT}    & \ding{55} & 50.53 & 44.19 & 54.66  & 58.33 & 50.49 & 25.67 & 65.07 & 59.02 & 54.14  & 22.22  & 47.33 & 40.30 & 46.26 & 21.95 & 37.61 & 48.25 & 39.83  & 48.33 & 47.47  \\
 & \ding{51} & 45.79 & 44.19 & 59.99  & 58.33 & 48.54 & 27.03 & 65.07 & 57.38 & 64.65  & 25.64  & 48.67 & 44.77 & 47.14 & 21.95 & 41.88 & 46.49 & 38.14  & 43.33 & \textbf{48.08}   \\
\multirow{2}{*}{LAVisH}   & \ding{55} & 71.90 & 61.27 & 56.47  & 58.79 & 71.84 & 27.87 & 43.68 & 63.13 & 70.99  & 41.53  & 35.62 & 57.37 & 66.75 & 21.91 & 43.92 & 69.38 & 28.76  & 44.91 & 54.88  \\
 & \ding{51} & 71.63 & 56.11 & 61.13  & 53.73 & 69.32 & 23.78 & 59.08 & 65.20 & 79.02  & 32.81  & 41.82 & 63.72 & 58.80 & 17.49 & 51.75 & 63.32 & 32.15  & 51.06 & \textbf{56.23}   \\ \midrule
VideoLLaMA 2    & \ding{55} & 89.47 & 67.44 & 64.00  & 40.00 & 84.95 & 67.57 & 55.02 & 74.59 & 87.97  & 44.44  & 61.33 & 52.24 & 70.48 & 51.22 & 59.83 & 62.28 & 49.15  & 68.33 & 66.85  \\
\multirow{2}{*}{MAVEN}    & \ding{55} & 88.64 & 62.50 & 85.71  & 33.33 & 92.31 & 66.67 & 87.23 & 75.76 & 85.71  & 89.29  & 65.62 & 55.56 & 71.43 & 44.44 & 45.83 & 61.90 & 54.55  & 85.71 & 72.92  \\
 & \ding{51} & 90.00 & 72.09 & 57.33  & 61.67 & 88.35 & 67.57 & 86.60 & 79.51 & 85.71  & 88.03  & 58.00 & 64.18 & 75.77 & 68.29 & 63.25 & 54.39 & 48.31  & 88.33 & \textbf{74.66}   \\
VITA  & \ding{55} & 91.45 & 48.28 & 62.67  & 35.00 & 74.03 & 38.89 & 15.74 & 46.19 & 63.16  & 45.30  & 34.34 & 37.29 & 47.22 & 27.27 & 77.78 & 28.95 & 27.27  & 49.50 & 48.66  \\
GPT 4o& \ding{55} & 63.78 & 46.51 & 69.33  & 16.67 & 75.00 & 68.92 & 56.52 & 75.21 & 44.36  & 87.18  & 42.95 & 50.75 & 56.83 & 39.02 & 13.68 & 63.16 & 30.51  & 71.67 & 56.06  \\
Qwen2.5-VL & \ding{55} & 88.11 & 57.14 & 83.78  & 13.33 & 77.94 & 72.97 & 46.38 & 68.91 & 57.14  & 76.07  & 36.24 & 53.73 & 56.76 & 60.98 & 11.97 & 76.99 & 25.42  & 63.33 & 57.07 \\ \bottomrule  
\end{tabular}
}
\end{table*}
MUSIC-AVQA \cite{li2022learning}, which contains training, validation, and testing splits with 31,927, 4,568, and 9,129 QA pairs, is developed by gathering questions for 9,288 musical performances. The questions are produced by a limited set of pre-defined templates. The videos, sourced from YouTube, include solo performances, ensembles of the same instruments, and ensembles of different instruments. This dataset consists of three tasks: audio QA, visual QA, and AVQA. The standard accuracy is used to evaluate model performance on the mentioned tasks. To comprehensively assess model robustness, we conduct rephrasing and splitting on the test split, expanding the question count from 9,129 to 211,572. Owing to the introduction of distribution shift, our proposed dataset provides three metrics: \textit{head accuracy, tail accuracy, and overall accuracy}, to evaluate models precisely. To reduce the test cost, we employ uniform sampling on the mentioned datasets at ratios of \(10\%\) and \(1\%\), yielding MUSIC-AVQA\textsubscript{s} and FortisAVQA\textsubscript{s}, respectively. The two sampled subsets are specifically used for evaluating large models, with detailed statistics provided in Table \ref{tab:test-comp}. Notably, the experiments for other baselines are conducted on the entire test split.  

\subsection{Implementation Details} In the collaborative debiasing module, we set the factors $\alpha$ and $\beta$ to 1$e$-3 and 5$e$-3 for optimization equilibrium, respectively. During the training stage,  we set the learning rate at an initial value of 2$e$-5, following a cosine decay schedule after a warm-up ratio of 0.03. The total number of training epochs is 2, using a batch size of 8 for training and 16 for evaluation. We employ the Adam optimizer and use gradient accumulation with a step size of 2 to stabilize training. The model is trained using DeepSpeed's ZeRO-3 optimization to efficiently handle memory usage.

The experiments for MAVEN are conducted using four NVIDIA Tesla A800 GPUs, while the other baseline experiments are performed on a single NVIDIA Tesla V100 GPU. Maven employs Mixtral 8×7B as the language foundation model.

\subsection{Baselines} We select 12 previous state-of-the-art small/large multimodal QA methods as baselines to verify the MAVEN effectiveness and investigate the robustness of these methods. Audio QA methods: FCNLSTM \cite{fayek2020temporal}, and CONVLSTM \cite{fayek2020temporal}. Visual QA methods: GRU \cite{antol2015vqa}, and MCAN \cite{yu2019deep}. Video QA methods: PSAC \cite{li2019beyond}, and HCRN \cite{le2020hierarchical}. AVQA methods: LAViT \cite{yun2021pano}, LAVisH \cite{lin2023vision}. Large multimodal models: VITA \cite{fu2024vita}, VideoLLaMa 2 \cite{cheng2024videollama}, Qwen2.5-VL \cite{bai2023qwen}, and GPT 4. Due to computing power limitations, we reevaluate LAVisH with a batch size of 2. For the plug-and-play experiments of MCCD on MCAN, GRU, HCRN, and PSAC, we set $\alpha$ and $\beta$ to 3$e$-3 and 1$e$-1, respectively. For experiments on CONVLSTM, LaViT, and LaVisH, $\alpha$ and $\beta$ were set to 1$e$-3 and 5$e$-3, respectively. For FCNLSTM, we set $\alpha$ and $\beta$ to 2$e$-3 and 5$e$-3, respectively. We conduct the experiments of large models except VideoLLaMA 2 in the zero-shot setting. The codes of baseline experiments are available at our released repository.
\begin{itemize}
    \item \textbf{FCNLSTM}\footnote{\url{https://github.com/facebookresearch/daqa}} integrates a fully convolutional network and LSTM to learn representations of audio and questions independently, and then projects their concatenated features into the answer space.
    \item \textbf{CONVLSTM} is a variant of FCNLSTM, incorporating five convolutional blocks identical to those in VGGNet to obtain variable-sized audio representations.
    \item \textbf{GRU} (referred to as ``deeper LSTM + Norm I'' in the original paper) serves as a simple baseline, employing VGGNet and LSTM to encode images and questions before mapping their concatenated features into the answer space.
    \item \textbf{MCAN}\footnote{\url{https://github.com/MILVLG/mcan-vqa}} is a deep modular co-attention network composed of cascaded modular co-attention layers, utilizing multi-head attention from Transformers for feature interaction.
    \item \textbf{PSAC}\footnote{\url{https://github.com/lixiangpengcs/PSAC}} leverages a positional self-attention block to model dependencies between question words and video frames separately, employing a co-attention mechanism for multimodal interaction.
    \item \textbf{HCRN}\footnote{\url{https://github.com/thaolmk54/hcrn-videoqa}} leverages a positional self-attention block to model dependencies between question words and video frames separately, employing a co-attention mechanism for multimodal interaction.
    \item \textbf{LAViT}\footnote{\url{https://github.com/hs-yn/PanoAVQA}} is a spatial AVQA framework that employs three distinct Transformer blocks to facilitate interactions between different input modalities.
    \item \textbf{LAVisH}\footnote{\url{https://github.com/GenjiB/LAVISH}} is a latent audio-visual hybrid adapter that adapts pre-trained Vision Transformers (ViTs) for audio-visual tasks by injecting a small number of trainable parameters into each layer of a frozen ViT.
    \item \textbf{VideoLLaMa 2}\footnote{\url{https://github.com/DAMO-NLP-SG/VideoLLaMA2}} is a video-based large language model featuring a custom-designed spatial-temporal convolution connector to capture intricate spatial and temporal video dynamics.
    \item \textbf{Qwen2.5-VL}\footnote{\url{https://huggingface.co/Qwen/Qwen2.5-VL-7B-Instruct}} is a multimodal large model that significantly enhances capabilities in visual recognition, precise object localization, robust document parsing, and long-video comprehension.
    \item \textbf{GPT 4o}\footnote{\url{https://chatgpt.com/}} is a large multimodal model capable of processing image and text inputs while generating text outputs, demonstrating human-level performance across various professional and academic benchmarks.
    \item \textbf{VITA}\footnote{\url{https://github.com/VITA-MLLM/VITA}} is the first open-source multimodal large language model, capable of processing video, image, text, and audio modalities while enabling interactive multimodal understanding.
\end{itemize}

\subsection{Comparison on MUSIC-AVQA} We conduct experiments on the MUSIC-AVQA test split to evaluate the effectiveness of the proposed method and compare its performance with existing baselines. The results are summarized in Table \ref{tab:exp-music}. All small models, except LAVisH, utilize ResNet-18/101 for visual feature encoding, whereas LAVisH leverages more powerful architectures such as ViT \cite{dosovitskiy2020image} and Swin Transformer\cite{liu2022swin} for visual representation learning. 

In the audio QA experiment, MAVEN achieves a new state-of-the-art accuracy of 72.79\%, outperforming the multimodal large model VideoLLaMA 2 by a significant margin of 3.15\%. Notably, large models do not always surpass small models in in-distribution scenarios; for instance, LAViT exceeds VideoLLaMA 2 by 1.22\%. The audio QA baselines do not show competitive results owing to the shallow multimodal interaction ability.

In the visual QA experiment, our method achieves the highest accuracy of 86.99\%, surpassing VideoLLaMA 2 by 4.88\%. We observe that visual and AVQA baselines consistently outperform other baselines. Similarly, in the AVQA experiment, MAVEN outperforms VideoLLaMA 2 by 0.96\%, achieving the best overall performance. Additionally, LAVisH’s incorporation of trainable parameters into robust visual encoders leads to superior results compared to the method relying on weaker encoders. 

Furthermore, AVQA methods like LAVisH outperform other approaches, likely due to the latter's inability to process audio, video, and text inputs simultaneously. In the zero-shot setting, all large models perform poorly, suggesting that their ability to process multiple modalities simultaneously requires further enhancement.

To evaluate the plug-and-play capability, we perform integration experiments with nine baseline models on the MUSIC-AVQA test split. In audio QA, MCCD enhances the performance of seven baselines, achieving a notable improvement of 6.52\% on HCRN. In visual QA, the strategy boosts five baselines, resulting in a significant gain of 8.46\% on HCRN. For AVQA, MCCD improves eight baselines, with a remarkable increase of 17.83\% on HCRN. The above results can demonstrate the efficacy of these methods and MCCD to some extent. However, MUSIC-AVQA lacks refined and precise evaluations due to its inherent shortcomings that are introduced in Section \ref{sec:intro}. \textit{Consequently, it is insufficient to evaluate these methods only on this dataset.} 
\begin{table*}
\caption{Ablation results (\%) on the test split of MUSIC-AVQA and our dataset. AQA and VQA denote audio QA, and visual QA, respectively. $D_{\mathrm{KL}}^{\#}$ is the distribution difference between the (\#) logit and the multimodality logit. MD: multifaceted debiasing. CG: cycle guidance.}
\label{tab:aba}
\centering

\begin{tabular}{c|cccc|cccccc}
\toprule
\multirow{2}{*}{\textbf{Setting}} & \multicolumn{4}{c|}{\textbf{MUSIC-AVQA}}  & \multicolumn{6}{c}{\textbf{FortisAVQA}}  \\ \cmidrule(lr){2-5} \cmidrule(lr){6-11}
& \textbf{AQA} & \textbf{VQA} & \textbf{AVQA}  & \textbf{All}   & \textbf{AQA} & \textbf{VQA} & \textbf{AVQA}  & \textbf{H}  & \textbf{T}  & \textbf{All}\\ \midrule
MAVEN  & \textbf{69.05}  & \textbf{84.55}  & \textbf{65.32} & \textbf{71.06} & \textbf{76.36} & \textbf{77.41}  & \textbf{63.81} & \textbf{72.14} & \textbf{65.33} & \textbf{69.90} \\
w/o $D_{\mathrm{KL}}^{\mathrm{q}}$  & 69.54       & 83.74     & 66.86 & 71.81 & 76.09    & 78.07     & 63.90 & 72.98 & 64.18 & 70.09 \\
w/o $D_{\mathrm{KL}}^{\mathrm{v}}$     & 69.05    & \textbf{84.96}     & \textbf{67.44}     & \textbf{72.35} & 75.82    & 77.25     & \textbf{65.21} & \textbf{73.33} & 64.76 & \textbf{70.51} \\
w/o $D_{\mathrm{KL}}^{\mathrm{a}}$    & 67.26    & 80.89     & 66.67     & 70.53 & 68.75    & 74.30     & 59.97 & 72.98 & 50.57 & 65.61 \\
w/o MD & \textbf{69.64}     & 83.33     & 65.70 & 71.06    & 75.00     & 76.76 & 64.07 & 72.77 & 63.81  & 69.62      \\
w/o CG & 68.45    & 86.18        & 66.09 & 71.81 & 74.46    & 77.09     & 63.72 & 71.58 & 65.04 & 69.43 \\ \bottomrule 
\end{tabular} 

\end{table*}

\subsection{Robustness Evaluation} 
We conduct experiments on the FortisAVQA test split to assess the robustness of both baseline models and MAVEN, as presented in Table \ref{tab:music-r}. Several crucial insights arise when combining the results from this table with those from Table \ref{tab:exp-music}.

First, simple multimodal fusion architectures exhibit notable robustness in smaller models. For instance, audio QA models such as CONVLSTM achieve competitive results, even surpassing other small models in tail accuracy on EXIST questions within the AVQA task. This occurs despite their relatively low in-distribution performance. In contrast, video QA and AVQA baseline models demonstrate strong performance on the MUSIC-AVQA test split but struggle on the FortisAVQA test split, particularly in the tail set, where LAViT achieves only 22.22\% accuracy on EXIST questions. This discrepancy suggests that their high performance on MUSIC-AVQA may stem from memorizing statistical regularities rather than genuine multimodality understanding.

Second, large models consistently exhibit greater robustness than small models. For example, while smaller models like LAViT and LAVisH perform comparably to the large model VideoLLaMA 2 on the MUSIC-AVQA test split, they suffer a substantial performance drop on the FortisAVQA test split. In contrast, VideoLLaMA 2 and MAVEN experience performance drops of only 5.71\% and 1.68\%, respectively. Notably, the three large models in the zero-shot setting even outperform their MUSIC-AVQA test split results, achieving performance levels comparable to smaller models trained with supervised learning. This highlights the superior robustness of large-scale multimodal models.

Third, MCCD, a plug-and-play debiasing strategy, effectively enhances the robustness of multimodal models. Specifically, MCCD improves performance for 7 out of 9 models on the FortisAVQA test split, mirroring its effect on the MUSIC-AVQA test split. For instance, it boosts MCAN and MAVEN by 1.69\% and 1.74\%, respectively.

In summary, the results presented in Tables \ref{tab:exp-music} and \ref{tab:music-r} confirm the high robustness of MAVEN and demonstrate the efficacy of MCCD as a plug-and-play debiasing strategy.

\subsection{Ablation Study} 
To verify the debiasing effectiveness of MCCD, we conduct extensive experiments on both the test split of MUSIC-AVQA and FortisAVQA. The results are shown in Table \ref{tab:aba}. Firstly, we validate the contribution of the component within multifaceted debiasing. It can be seen that removing the component will lead to an overall performance improvement in some aspects of MUSIC-AVQA while resulting in a significant decrease in the tail split of FortisAVQA or the out-of-distribution scenario. This observation strongly supports the debiasing efficacy of these components. It should be noted that removing the audio logit regularization leads to a significant performance decreasement in both datasets. This indicates that audio distribution constraints are crucial in improving robustness, particularly in audio-centric tasks. Secondly, we verify the overall contribution of the multifaceted debiasing. It can be seen that the performance decreasement of 1.52\% occurs in the tail split. Finally, we validate the contribution of cycle guidance. We see that this model variant obtains the performance improvement on the MUSIC-AVQA dataset or the in-distribution scenario. However, there was a noticeable performance degradation in each task of FortisAVQA. In summary, each component plays a distinctive role in the debiasing process, which is further demonstrated by the performance degradation on the tail split.

\subsection{Sensitivity Analysis} To examine the impact of hyperparameters $\alpha$ and $\beta$ on the robustness of MAVEN multimodal reasoning, we conduct a sensitivity analysis by varying one parameter while keeping the other fixed.
\begin{figure}[t]
    \centering
    \subfigure{\includegraphics[width=0.48\columnwidth]{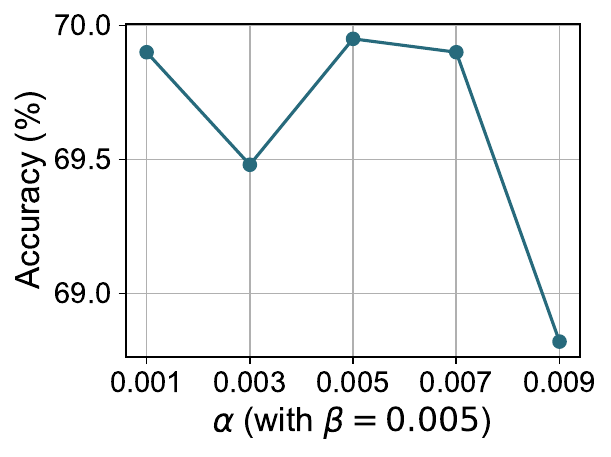}} 
    \hspace{1mm} % 水平间距，根据需要调整
    \subfigure{\includegraphics[width=0.48\columnwidth]{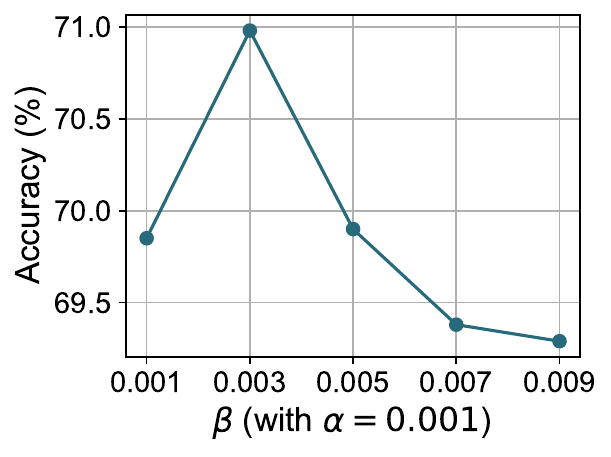}}
    \caption{Senstivity analysis of discrepancy enlargement and cycle constraint in Equation \eqref{eq:dist} and \eqref{eq:cyc}.}
    \label{fig:sens}
\end{figure}

As shown in Figure \ref{fig:sens}, we fix $\alpha$ = 0.001 and analyze the effect of $\beta$, which controls the regularization strength of the cycle constraint loss. Our method achieves the highest accuracy (70.98\%) at $\beta$ = 0.003, but further increasing $\beta$ leads to a decline in accuracy. This suggests that while an appropriate $\beta$ can effectively reinforce the consistency of three unimodal logits, an excessively large $\beta$ may impose excessive regularization, limiting the model to generalize effectively.

Similarly, we fix $\beta$ = 0.005 and evaluate the effect of $\alpha$, which regulates the loss enforcing the divergence between unimodal and multimodal logits. The results exhibit fluctuations in accuracy, with the best performance observed at $\alpha$ = 0.001 and $\alpha$ = 0.005, while larger $\alpha$ values result in a significant accuracy drop. This indicates that while $\alpha$ is crucial for ensuring a meaningful distinction between unimodal and multimodal representations, an overly strong divergence constraint may hinder the model from generating answers depending on beneficial biases.

Overall, these results highlight the importance of carefully balancing $\alpha$ and $\beta$ to optimize the model bias mitigation strategy and maximize performance in multimodal question answering.

\subsection{Qualitative Analysis} 
\begin{figure*}[tbp]
    \centering  %图片全局居中
    \includegraphics[width=\linewidth]{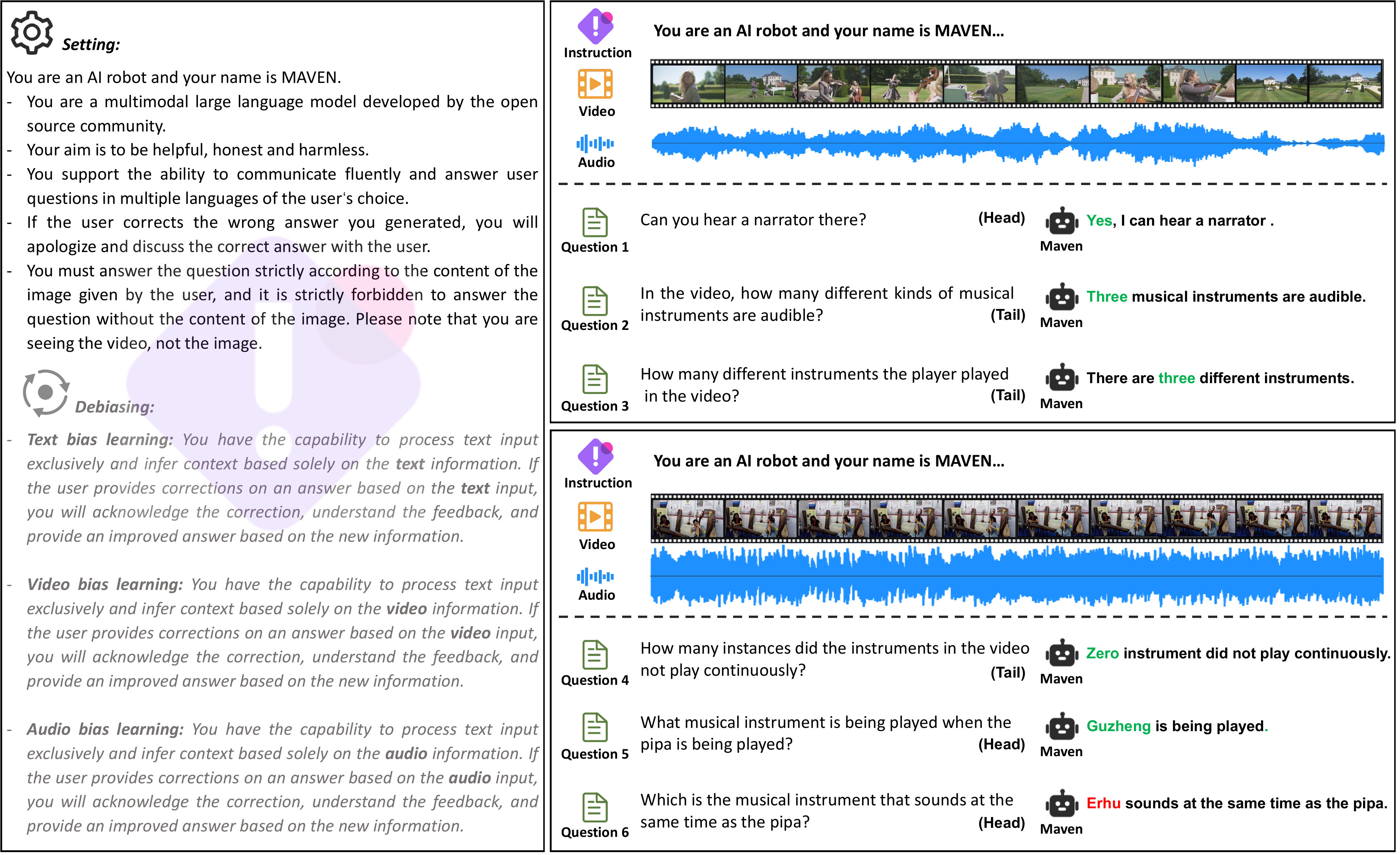}
    \caption{Qualitative analysis of FortisAVQA and MAVEN. The left panel presents the instruction of multimodal comprehension and modality-specific bias learning. The right panel shows the answer generation of MAVEN for the samples of FortisAVQA.}
    \label{fig:case}
\end{figure*}
We conduct a qualitative analysis on FortisAVQA and MAVEN to show the robustness evaluation and debiasing capability in Fig. \ref{fig:case}. In the right panel, we observe that FortisAVQA effectively assesses model performance in both in-distribution and out-of-distribution scenarios. To enhance diversity and realism, we rephrase the questions multiple times, as exemplified by Questions 1 and 3. This demonstrates the dataset's comprehensive and fine-grained robustness. In the left panel, we present multimodal fusion and modality-specific bias-capturing instructions. The former guides the model to generate responses by integrating information from all input modalities, while the latter prompts the model to make predictions based on a single modality. The identified biases are then mitigated using the MCCD strategy. To evaluate the robustness of MAVEN, we select six questions across two videos and audio samples. In the top-right panel, our method consistently generates accurate responses for semantically equivalent tail questions (e.g., Question 2 and 3). However, we observe that MAVEN does not always produce accurate responses for head questions with different phrasings. This finding underscores the need for more robust multimodal models capable of making accurate predictions regardless of variations in question formulation.

\section{Conclusion and Future Work}
We are the first to explore bias learning in the AVQA task through both dataset evaluation and model design. To this end, we introduce FortisAVQA, a new dataset that assesses model performance across head, tail, and overall samples, providing a precise measure of robustness. Additionally, we propose a robust generative architecture leveraging Multifaceted Cycle Collaborative Debiasing (MCCD) to mitigate bias learning. Extensive experiments validate the effectiveness of our approach and highlight the plug-and-play debiasing capability of MCCD. Furthermore, our re-evaluation of previous multimodal QA models on FortisAVQA reveals significant robustness issues. In future work, we will explore mitigating biased predictions through test-time computation and reinforcement learning.

\bibliographystyle{IEEEtran}
\bibliography{reference}

\begin{thebibliography}{10}
\providecommand{\url}[1]{#1}
\csname url@samestyle\endcsname
\providecommand{\newblock}{\relax}
\providecommand{\bibinfo}[2]{#2}
\providecommand{\BIBentrySTDinterwordspacing}{\spaceskip=0pt\relax}
\providecommand{\BIBentryALTinterwordstretchfactor}{4}
\providecommand{\BIBentryALTinterwordspacing}{\spaceskip=\fontdimen2\font plus
\BIBentryALTinterwordstretchfactor\fontdimen3\font minus \fontdimen4\font\relax}
\providecommand{\BIBforeignlanguage}[2]{{%
\expandafter\ifx\csname l@#1\endcsname\relax
\typeout{** WARNING: IEEEtran.bst: No hyphenation pattern has been}%
\typeout{** loaded for the language `#1'. Using the pattern for}%
\typeout{** the default language instead.}%
\else
\language=\csname l@#1\endcsname
\fi
#2}}
\providecommand{\BIBdecl}{\relax}
\BIBdecl

\bibitem{lin2023vision}
Y.-B. Lin, Y.-L. Sung, J.~Lei, M.~Bansal, and G.~Bertasius, ``Vision transformers are parameter-efficient audio-visual learners,'' in \emph{CVPR}, 2023, pp. 2299--2309.

\bibitem{ma2024diagram}
J.~Ma, J.~Liu, Q.~Chai, P.~Wang, and J.~Tao, ``Diagram perception networks for textbook question answering via joint optimization,'' \emph{IJCV}, vol. 132, no.~5, pp. 1578--1591, 2024.

\bibitem{ma2022weakly}
J.~Ma, Q.~Chai, J.~Huang, J.~Liu, Y.~You, and Q.~Zheng, ``Weakly supervised learning for textbook question answering,'' \emph{IEEE TIP}, vol.~31, pp. 7378--7388, 2022.

\bibitem{alamri2019audio}
H.~Alamri, V.~Cartillier, A.~Das, J.~Wang, A.~Cherian, I.~Essa, D.~Batra, T.~K. Marks, C.~Hori, P.~Anderson \emph{et~al.}, ``Audio visual scene-aware dialog,'' in \emph{CVPR}, 2019, pp. 7558--7567.

\bibitem{yang2022avqa}
P.~Yang, X.~Wang, X.~Duan, H.~Chen, R.~Hou, C.~Jin, and W.~Zhu, ``{AVQA}: A dataset for audio-visual question answering on videos,'' in \emph{ACM MM}, 2022, pp. 3480--3491.

\bibitem{li2022learning}
G.~Li, Y.~Wei, Y.~Tian, C.~Xu, J.-R. Wen, and D.~Hu, ``Learning to answer questions in dynamic audio-visual scenarios,'' in \emph{CVPR}, 2022, pp. 19\,108--19\,118.

\bibitem{yun2021pano}
H.~Yun, Y.~Yu, W.~Yang, K.~Lee, and G.~Kim, ``{Pano-AVQA}: Grounded audio-visual question answering on 360${}^\circ$ videos,'' in \emph{CVPR}, 2021, pp. 2031--2041.

\bibitem{wen2021debiased}
Z.~Wen, G.~Xu, M.~Tan, Q.~Wu, and Q.~Wu, ``Debiased visual question answering from feature and sample perspectives,'' in \emph{NeurIPS}, 2021, pp. 3784--3796.

\bibitem{vatsa2024adventures}
M.~Vatsa, A.~Jain, and R.~Singh, ``Adventures of trustworthy vision-language models: A survey,'' in \emph{AAAI}, 2024, pp. 22\,650--22\,658.

\bibitem{hall2024visogender}
S.~M. Hall, F.~Gon{\c{c}}alves~Abrantes, H.~Zhu, G.~Sodunke, A.~Shtedritski, and H.~R. Kirk, ``Visogender: A dataset for benchmarking gender bias in image-text pronoun resolution,'' in \emph{NeurIPS}, 2024.

\bibitem{torralba2011}
A.~Torralba and A.~A. Efros, ``Unbiased look at dataset bias,'' in \emph{CVPR}, 2011, pp. 1521--1528.

\bibitem{li2023multi}
Y.~Li, B.~Hu, F.~Zhang, Y.~Yu, J.~Liu, Y.~Chen, and J.~Xu, ``A multi-modal debiasing model with dynamical constraint for robust visual question answering,'' in \emph{Findings of ACL}, 2023, pp. 5032--5045.

\bibitem{ravichander2023and}
A.~Ravichander, J.~Stacey, and M.~Rei, ``When and why does bias mitigation work?'' in \emph{Findings of EMNLP}, 2023, pp. 9233--9247.

\bibitem{agrawal2018don}
A.~Agrawal, D.~Batra, D.~Parikh, and A.~Kembhavi, ``Don't just assume; look and answer: Overcoming priors for visual question answering,'' in \emph{CVPR}, 2018, pp. 4971--4980.

\bibitem{kervadec2021roses}
C.~Kervadec, G.~Antipov, M.~Baccouche, and C.~Wolf, ``Roses are red, violets are blue... but should {VQA} expect them to?'' in \emph{CVPR}, 2021, pp. 2776--2785.

\bibitem{niu2021introspective}
Y.~Niu and H.~Zhang, ``Introspective distillation for robust question answering,'' in \emph{NeurIPS}, 2021, pp. 16\,292--16\,304.

\bibitem{xue2024integrating}
D.~Xue, S.~Qian, and C.~Xu, ``Integrating neural-symbolic reasoning with variational causal inference network for explanatory visual question answering,'' \emph{IEEE TPAMI}, vol.~46, no.~12, pp. 7893 -- 7908, 2024.

\bibitem{liu2024tackling}
X.~Liu, Z.~Dong, and P.~Zhang, ``Tackling data bias in music-avqa: Crafting a balanced dataset for unbiased question-answering,'' in \emph{WACV}, 2024, pp. 4478--4487.

\bibitem{zhangZX24}
X.~Zhang, F.~Zhang, and C.~Xu, ``Next-ood: Overcoming dual multiple-choice {VQA} biases,'' \emph{IEEE TPAMI}, vol.~46, no.~4, pp. 1913--1931, 2024.

\bibitem{girdhar2023}
R.~Girdhar, A.~El-Nouby, Z.~Liu, M.~Singh, K.~V. Alwala, A.~Joulin, and I.~Misra, ``{ImageBind}: One embedding space to bind them all,'' in \emph{CVPR}, 2023, pp. 15\,180--15\,190.

\bibitem{zhangLB23}
H.~Zhang, X.~Li, and L.~Bing, ``{Video-LLaMA}: An instruction-tuned audio-visual language model for video understanding,'' in \emph{EMNLP (Demos)}, 2023, pp. 543--553.

\bibitem{ma2024look}
J.~Ma, M.~Hu, P.~Wang, W.~Sun, L.~Song, H.~Pei, J.~Liu, and Y.~Du, ``Look, listen, and answer: Overcoming biases for audio-visual question answering,'' \emph{NeurIPS}, 2024.

\bibitem{oliveira2024split}
R.~I. Oliveira, P.~Orenstein, T.~Ramos, and J.~V. Romano, ``Split conformal prediction and non-exchangeable data,'' \emph{JMLR}, vol.~25, no. 225, pp. 1--38, 2024.

\bibitem{lyu2023macaw}
C.~Lyu, M.~Wu, L.~Wang, X.~Huang, B.~Liu, Z.~Du, S.~Shi, and Z.~Tu, ``Macaw-llm: Multi-modal language modeling with image, audio, video, and text integration,'' \emph{arXiv preprint arXiv:2306.09093}, 2023.

\bibitem{guo2025deepseek}
D.~Guo, D.~Yang, H.~Zhang, J.~Song, R.~Zhang, R.~Xu, Q.~Zhu, S.~Ma, P.~Wang, X.~Bi \emph{et~al.}, ``Deepseek-r1: Incentivizing reasoning capability in llms via reinforcement learning,'' \emph{arXiv preprint arXiv:2501.12948}, 2025.

\bibitem{bai2023qwen}
J.~Bai, S.~Bai, S.~Yang, S.~Wang, S.~Tan, P.~Wang, J.~Lin, C.~Zhou, and J.~Zhou, ``Qwen-vl: A frontier large vision-language model with versatile abilities,'' \emph{arXiv preprint arXiv:2308.12966}, 2023.

\bibitem{li2021align}
J.~Li, R.~Selvaraju, A.~Gotmare, S.~Joty, C.~Xiong, and S.~C.~H. Hoi, ``Align before fuse: Vision and language representation learning with momentum distillation,'' in \emph{NeurIPS}, 2021, pp. 9694--9705.

\bibitem{radford2021learning}
A.~Radford, J.~W. Kim, C.~Hallacy, A.~Ramesh, G.~Goh, S.~Agarwal, G.~Sastry, A.~Askell, P.~Mishkin, J.~Clark \emph{et~al.}, ``Learning transferable visual models from natural language supervision,'' in \emph{ICML}, 2021, pp. 8748--8763.

\bibitem{li2022blip}
J.~Li, D.~Li, C.~Xiong, and S.~Hoi, ``Blip: Bootstrapping language-image pre-training for unified vision-language understanding and generation,'' in \emph{ICML}.\hskip 1em plus 0.5em minus 0.4em\relax PMLR, 2022, pp. 12\,888--12\,900.

\bibitem{jia2021scaling}
C.~Jia, Y.~Yang, Y.~Xia, Y.-T. Chen, Z.~Parekh, H.~Pham, Q.~Le, Y.-H. Sung, Z.~Li, and T.~Duerig, ``Scaling up visual and vision-language representation learning with noisy text supervision,'' in \emph{ICML}, 2021, pp. 4904--4916.

\bibitem{ma2021multitask}
J.~Ma, J.~Liu, Q.~Lin, B.~Wu, Y.~Wang, and Y.~You, ``Multitask learning for visual question answering,'' \emph{IEEE TNNLS}, vol.~34, no.~3, pp. 1380--1394, 2021.

\bibitem{yu2022coca}
J.~Yu, Z.~Wang, V.~Vasudevan, L.~Yeung, M.~Seyedhosseini, and Y.~Wu, ``Coca: Contrastive captioners are image-text foundation models,'' \emph{TMLR}, vol. 2022, 2022.

\bibitem{li2023blip}
J.~Li, D.~Li, S.~Savarese, and S.~Hoi, ``Blip-2: Bootstrapping language-image pre-training with frozen image encoders and large language models,'' in \emph{ICML}, 2023, pp. 19\,730--19\,742.

\bibitem{jian2024}
Y.~Jian, C.~Gao, and S.~Vosoughi, ``Bootstrapping vision-language learning with decoupled language pre-training,'' in \emph{NeurIPS}, 2024.

\bibitem{wu2024building}
Z.~Wu, Z.~Weng, W.~Peng, X.~Yang, A.~Li, L.~S. Davis, and Y.-G. Jiang, ``Building an open-vocabulary video clip model with better architectures, optimization and data,'' \emph{IEEE TPAMI}, 2024.

\bibitem{dai2023instructblip}
W.~Dai, J.~Li, D.~Li, A.~M.~H. Tiong, J.~Zhao, W.~Wang, B.~Li, P.~Fung, and S.~Hoi, ``Instructblip: Towards general-purpose vision-language models with instruction tuning,'' \emph{arXiv preprint arXiv:2305.06500}, 2023.

\bibitem{zhuminigpt}
D.~Zhu, J.~Chen, X.~Shen, X.~Li, and M.~Elhoseiny, ``Minigpt-4: Enhancing vision-language understanding with advanced large language models,'' in \emph{ICLR}, 2023.

\bibitem{cheng2024videollama}
Z.~Cheng, S.~Leng, H.~Zhang, Y.~Xin, X.~Li, G.~Chen, Y.~Zhu, W.~Zhang, Z.~Luo, D.~Zhao \emph{et~al.}, ``Videollama 2: Advancing spatial-temporal modeling and audio understanding in video-llms,'' \emph{arXiv preprint arXiv:2406.07476}, 2024.

\bibitem{fu2024vita}
C.~Fu, H.~Lin, Z.~Long, Y.~Shen, M.~Zhao, Y.~Zhang, S.~Dong, X.~Wang, D.~Yin, L.~Ma \emph{et~al.}, ``Vita: Towards open-source interactive omni multimodal llm,'' \emph{arXiv preprint arXiv:2408.05211}, 2024.

\bibitem{xu2023multiinstruct}
Z.~Xu, Y.~Shen, and L.~Huang, ``Multiinstruct: Improving multi-modal zero-shot learning via instruction tuning,'' in \emph{ACL}, 2023, pp. 11\,445--11\,465.

\bibitem{zhang2024video}
Y.~Zhang, J.~Wu, W.~Li, B.~Li, Z.~Ma, Z.~Liu, and C.~Li, ``Video instruction tuning with synthetic data,'' \emph{arXiv preprint arXiv:2410.02713}, 2024.

\bibitem{su2023pandagpt}
Y.~Su, T.~Lan, H.~Li, J.~Xu, Y.~Wang, and D.~Cai, ``Pandagpt: One model to instruction-follow them all,'' in \emph{Proceedings of the 1st Workshop on Taming Large Language Models: Controllability in the era of Interactive Assistants!}, 2023, pp. 11--23.

\bibitem{goyal2019making}
Y.~Goyal, T.~Khot, A.~Agrawal, D.~Summers-Stay, D.~Batra, and D.~Parikh, ``Making the v in {VQA} matter: Elevating the role of image understanding in visual question answering,'' \emph{IJCV}, vol. 127, pp. 398--414, 2019.

\bibitem{hudson2019gqa}
D.~A. Hudson and C.~D. Manning, ``{GQA}: A new dataset for real-world visual reasoning and compositional question answering,'' in \emph{CVPR}, 2019, pp. 6700--6709.

\bibitem{yang2022}
A.~Yang, A.~Miech, J.~Sivic, I.~Laptev, and C.~Schmid, ``Zero-shot video question answering via frozen bidirectional language models,'' in \emph{NeurIPS}, S.~Koyejo, S.~Mohamed, A.~Agarwal, D.~Belgrave, K.~Cho, and A.~Oh, Eds., 2022, pp. 124--141.

\bibitem{ma2023robust}
J.~Ma, P.~Wang, D.~Kong, Z.~Wang, J.~Liu, H.~Pei, and J.~Zhao, ``Robust visual question answering: Datasets, methods, and future challenges,'' \emph{IEEE TPAMI}, vol.~46, no.~8, pp. 5575--5594, 2024.

\bibitem{kervadec2021}
C.~Kervadec, C.~Wolf, G.~Antipov, M.~Baccouche, and M.~Nadri, ``Supervising the transfer of reasoning patterns in vqa,'' in \emph{NeurIPS}, vol.~34, 2021, pp. 18\,256--18\,267.

\bibitem{ramakrishnanAL18}
S.~Ramakrishnan, A.~Agrawal, and S.~Lee, ``Overcoming language priors in visual question answering with adversarial regularization,'' in \emph{NeurIPS}, 2018, pp. 1548--1558.

\bibitem{zheng2023large}
C.~Zheng, H.~Zhou, F.~Meng, J.~Zhou, and M.~Huang, ``Large language models are not robust multiple choice selectors,'' in \emph{ICLR}, 2023.

\bibitem{ko2020look}
M.~Ko, J.~Lee, H.~Kim, G.~Kim, and J.~Kang, ``Look at the first sentence: Position bias in question answering,'' in \emph{EMNLP}, 2020, pp. 1109--1121.

\bibitem{dancette2021beyond}
C.~Dancette, R.~Cadene, D.~Teney, and M.~Cord, ``Beyond question-based biases: Assessing multimodal shortcut learning in visual question answering,'' in \emph{ICCV}, 2021, pp. 1574--1583.

\bibitem{harrell1996}
F.~E. Harrell~Jr, K.~L. Lee, and D.~B. Mark, ``Multivariable prognostic models: Issues in developing models, evaluating assumptions and adequacy, and measuring and reducing errors,'' \emph{Statistics in medicine}, vol.~15, no.~4, pp. 361--387, 1996.

\bibitem{sheng2021human}
S.~Sheng, A.~Singh, V.~Goswami, J.~Magana, T.~Thrush, W.~Galuba, D.~Parikh, and D.~Kiela, ``Human-adversarial visual question answering,'' in \emph{NeurIPS}, 2021, pp. 20\,346--20\,359.

\bibitem{li2021adversarial}
L.~Li, J.~Lei, Z.~Gan, and J.~Liu, ``Adversarial {VQA}: A new benchmark for evaluating the robustness of {VQA} models,'' in \emph{ICCV}, 2021, pp. 2042--2051.

\bibitem{xu2023counterfactual}
W.~Xu, Q.~Liu, S.~Wu, and L.~Wang, ``Counterfactual debiasing for fact verification,'' in \emph{ACL}, 2023, pp. 6777--6789.

\bibitem{TsirigotisMR0C23}
C.~Tsirigotis, J.~Monteiro, P.~Rodr{\'{\i}}guez, D.~V{\'{a}}zquez, and A.~C. Courville, ``Group robust classification without any group information,'' in \emph{NeurIPS}, 2023.

\bibitem{esiobu2023robbie}
D.~Esiobu, X.~Tan, S.~Hosseini, M.~Ung, Y.~Zhang, J.~Fernandes, J.~Dwivedi-Yu, E.~Presani, A.~Williams, and E.~Smith, ``Robbie: Robust bias evaluation of large generative language models,'' in \emph{EMNLP}, 2023, pp. 3764--3814.

\bibitem{Cho0RK23}
J.~Cho, D.~Kim, H.~Ryu, and I.~S. Kweon, ``Generative bias for robust visual question answering,'' in \emph{CVPR}, 2023, pp. 11\,681--11\,690.

\bibitem{ma2023adaptive}
J.~Ma, P.~Wang, Z.~Wang, D.~Kong, M.~Hu, T.~Han, and J.~Liu, ``Adaptive loose optimization for robust question answering,'' \emph{arXiv preprint arXiv:2305.03971}, 2023.

\bibitem{abbasnejad2020counterfactual}
E.~Abbasnejad, D.~Teney, A.~Parvaneh, J.~Shi, and A.~v.~d. Hengel, ``Counterfactual vision and language learning,'' in \emph{CVPR}, 2020, pp. 10\,044--10\,054.

\bibitem{chen2022rethinking}
L.~Chen, Y.~Zheng, and J.~Xiao, ``Rethinking data augmentation for robust visual question answering,'' in \emph{ECCV}, 2022, pp. 95--112.

\bibitem{teney2021unshuffling}
D.~Teney, E.~Abbasnejad, and A.~van~den Hengel, ``Unshuffling data for improved generalization in visual question answering,'' in \emph{ICCV}, 2021, pp. 1417--1427.

\bibitem{liang2020learning}
Z.~Liang, W.~Jiang, H.~Hu, and J.~Zhu, ``Learning to contrast the counterfactual samples for robust visual question answering,'' in \emph{EMNLP}, 2020, pp. 3285--3292.

\bibitem{zhu2021overcoming}
X.~Zhu, Z.~Mao, C.~Liu, P.~Zhang, B.~Wang, and Y.~Zhang, ``Overcoming language priors with self-supervised learning for visual question answering,'' in \emph{IJCAI}, 2021, pp. 1083--1089.

\bibitem{si2022towards}
Q.~Si, Y.~Liu, F.~Meng, Z.~Lin, P.~Fu, Y.~Cao, W.~Wang, and J.~Zhou, ``Towards robust visual question answering: Making the most of biased samples via contrastive learning,'' in \emph{Findings of EMNLP}, 2022, pp. 6650--6662.

\bibitem{jing2020overcoming}
C.~Jing, Y.~Wu, X.~Zhang, Y.~Jia, and Q.~Wu, ``Overcoming language priors in {VQA} via decomposed linguistic representations,'' in \emph{AAAI}, 2020, pp. 11\,181--11\,188.

\bibitem{shrestha2020negative}
R.~Shrestha, K.~Kafle, and C.~Kanan, ``A negative case analysis of visual grounding methods for {VQA},'' in \emph{ACL}, 2020, pp. 8172--8181.

\bibitem{gat2020removing}
I.~Gat, I.~Schwartz, A.~Schwing, and T.~Hazan, ``Removing bias in multi-modal classifiers: Regularization by maximizing functional entropies,'' in \emph{NeurIPS}, 2020, pp. 3197--3208.

\bibitem{lao2023coca}
M.~Lao, N.~Pu, Y.~Liu, K.~He, E.~M. Bakker, and M.~S. Lew, ``{COCA}: Collaborative causal regularization for audio-visual question answering,'' in \emph{AAAI}, 2023, pp. 12\,995--13\,003.

\bibitem{lu2024unified}
J.~Lu, C.~Clark, S.~Lee, Z.~Zhang, S.~Khosla, R.~Marten, D.~Hoiem, and A.~Kembhavi, ``Unified-io 2: Scaling autoregressive multimodal models with vision language audio and action,'' in \emph{CVPR}, 2024, pp. 26\,439--26\,455.

\bibitem{fayek2020temporal}
H.~M. Fayek and J.~Johnson, ``Temporal reasoning via audio question answering,'' \emph{TASLP}, vol.~28, pp. 2283--2294, 2020.

\bibitem{antol2015vqa}
S.~Antol, A.~Agrawal, J.~Lu, M.~Mitchell, D.~Batra, C.~L. Zitnick, and D.~Parikh, ``{VQA}: Visual question answering,'' in \emph{ICCV}, 2015, pp. 2425--2433.

\bibitem{yu2019deep}
Z.~Yu, J.~Yu, Y.~Cui, D.~Tao, and Q.~Tian, ``Deep modular co-attention networks for visual question answering,'' in \emph{CVPR}, 2019, pp. 6281--6290.

\bibitem{li2019beyond}
X.~Li, J.~Song, L.~Gao, X.~Liu, W.~Huang, X.~He, and C.~Gan, ``Beyond {RNNs}: Positional self-attention with co-attention for video question answering,'' in \emph{AAAI}, 2019, pp. 8658--8665.

\bibitem{le2020hierarchical}
T.~M. Le, V.~Le, S.~Venkatesh, and T.~Tran, ``Hierarchical conditional relation networks for video question answering,'' in \emph{CVPR}, 2020, pp. 9972--9981.

\bibitem{dosovitskiy2020image}
A.~Dosovitskiy, L.~Beyer, A.~Kolesnikov, D.~Weissenborn, X.~Zhai, T.~Unterthiner, M.~Dehghani, M.~Minderer, G.~Heigold, S.~Gelly \emph{et~al.}, ``An image is worth 16x16 words: Transformers for image recognition at scale,'' in \emph{ICLR}, 2020.

\bibitem{liu2022swin}
Z.~Liu, H.~Hu, Y.~Lin, Z.~Yao, Z.~Xie, Y.~Wei, J.~Ning, Y.~Cao, Z.~Zhang, L.~Dong \emph{et~al.}, ``Swin transformer v2: Scaling up capacity and resolution,'' in \emph{CVPR}, 2022, pp. 12\,009--12\,019.

\end{thebibliography}

\ifCLASSOPTIONcaptionsoff
  \newpage
\fi
%\newpage
\vspace{-1cm}

\begin{IEEEbiography}[{\includegraphics[width=1in,height=1.25in,clip,keepaspectratio]{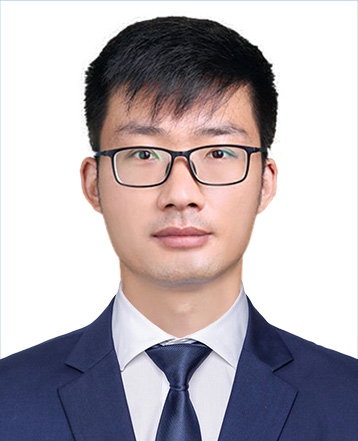}}]{Jie Ma}	
(Member, IEEE) is an Assistant Professor at the School of Cyber Science and Engineering, Xi’an Jiaotong University, and a member of MOE KLINNS Lab. He received joint Ph.D. training at the National University of Singapore from April 2021 to April 2022 and earned his Ph.D. from Xi’an Jiaotong University in July 2022. His research focuses on multimodal data mining and content understanding, natural language processing, and intelligent education, with particular emphasis on textbook question answering, reasoning with large models on knowledge graphs, and robust multimodal question answering. He has been recognized by the China Association for Science and Technology and the Shaanxi Association for Science and Technology through their Youth Talent Support Programs. He has led multiple research initiatives, including the NSFC Youth Project, as well as industry collaborations with China Mobile and other enterprises. He serves as a guest editor for Information Fusion, a leading international journal, and as an area chair for ICML and CCKS. Additionally, he is a member of the IEEE Knowledge Engineering Standards Committee and an executive committee member of several prestigious technical societies, including CCF, CAAI, and CIPS. He has authored over 20 papers in top-tier international journals and conferences such as IEEE TPAMI, IJCV, NeurIPS, and ICML. He is also an active reviewer for top journals like IEEE TIP and TNNLS and serves on the program committees of top conferences, such as NeurIPS and ICLR. For more information, please visit \url{https://dr-majie.github.io/}.
\end{IEEEbiography}

\begin{IEEEbiography}[{\includegraphics[width=1in,height=1.25in,clip,keepaspectratio]{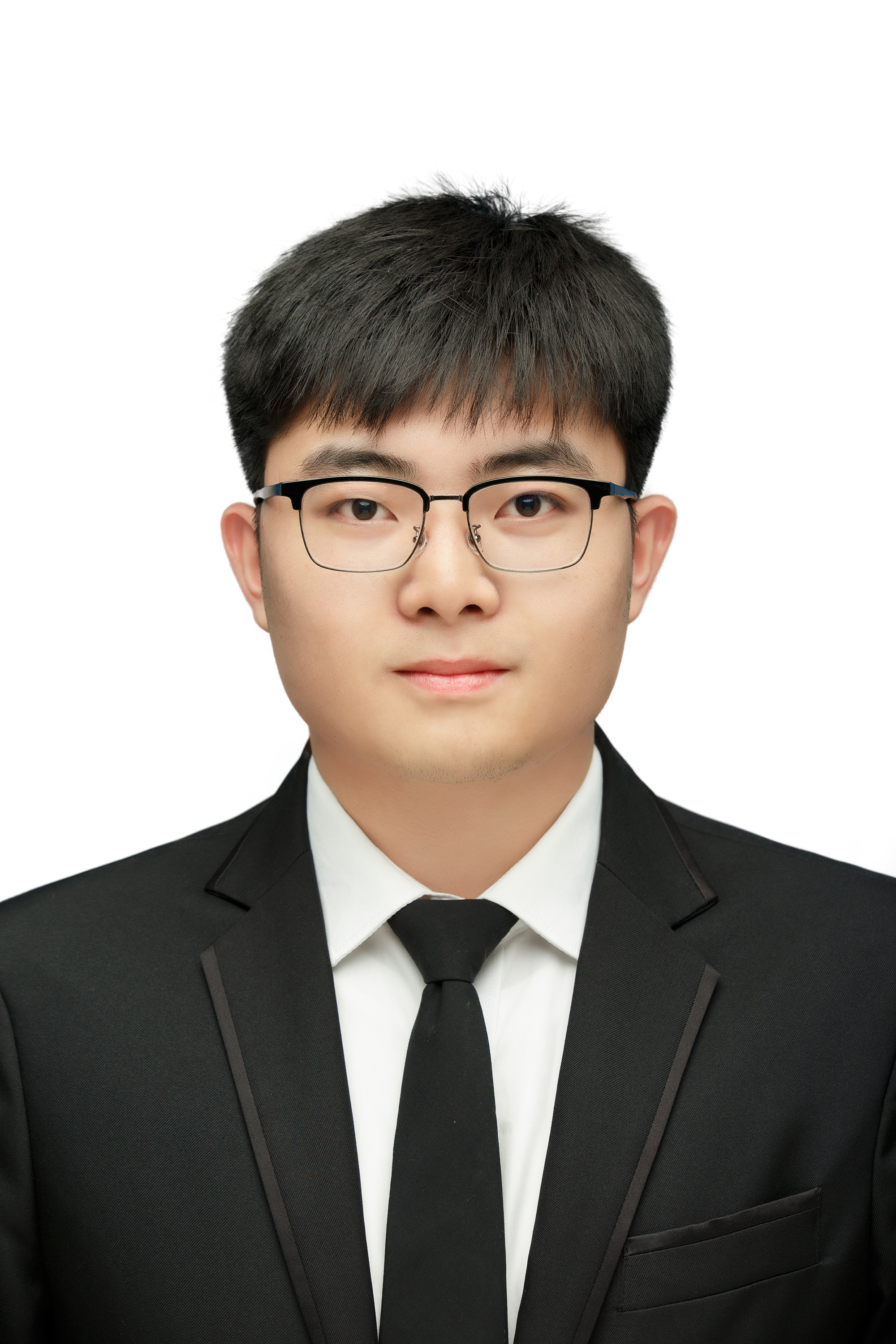}}]{Zhitao Gao}
is currently pursuing a bachelor's degree in Computer Science and Technology at Xi'an Jiaotong University. His research interests focus on natural language processing and multimodal learning.
\end{IEEEbiography}

\begin{IEEEbiography}[{\includegraphics[width=1in,height=1.25in,clip,keepaspectratio]{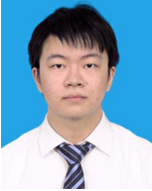}}]{Qi Chai}
 received his B.E. and M.S. degrees from Xi’an
 Jiaotong University, Xi’an, China in 2021 and 2024. He is currently a research assistant at the Information Hub, Hong Kong University of Science and Technology (Guangzhou), Guangzhou, China. His research interests include multi-modal question answering, image processing, and natural language processing.
\end{IEEEbiography}
\vspace{-0.5cm}

\begin{IEEEbiography}[{\includegraphics[width=1in,height=1.25in,clip,keepaspectratio]{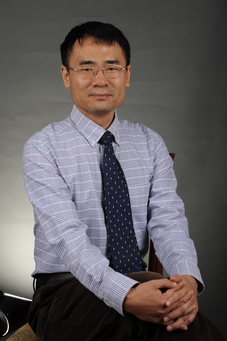}}]{Jun Liu}
(Senior Member, IEEE) received a B.S. in computer science and technology in 1995 and a Ph.D. degree in systems engineering in 2004, both from Xi’an Jiaotong University, China. He is currently a Professor with the Department of Computer Science at Xi’an Jiaotong University. He has authored more than 150 research papers in various journals and conference proceedings. He has won the best paper awards in IEEE ISSRE 2016 and IEEE ICBK 2016. His research interests include NLP, CV, and Smart Education. Dr. Liu currently has served as an associate editor of IEEE TNNLS and Information Fusion and has served as a guest editor for many technical journals, such as WWWJ and IEEE SYSTEMS JOURNAL. He also acted as a conference/workshop/track chair at numerous conferences. For more information, please visit \url{https://gr.xjtu.edu.cn/web/liukeen/1}.
\end{IEEEbiography}
\vspace{-0.5cm}

\begin{IEEEbiography}[{\includegraphics[width=1in,height=1.25in,clip,keepaspectratio]{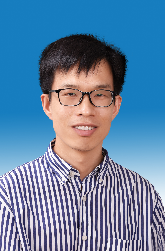}}]{Pinghui Wang}
(Senior Member, IEEE) is currently a professor with the MOE Key Laboratory for Intelligent Networks and Network Security, Xi’an Jiaotong University, Xi’an, China, and also with the Shenzhen Research Institute, Xi’an Jiaotong University, Shenzhen, China. His research interests include internet traffic measurement and modeling, large language models, abnormal detection, and online social network measurement.
\end{IEEEbiography}
\vspace{-0.5cm}

\begin{IEEEbiography}[{\includegraphics[width=1in,height=1.25in,clip,keepaspectratio]{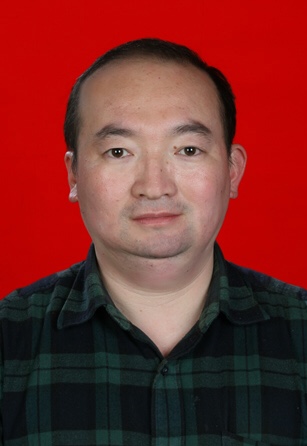}}]{Jing Tao}
received his B.S. and M.S. degrees in Automatic Control from Xi’an Jiaotong University, Xi’an, China, in 2001 and 2006, respectively. He is currently a faculty member at Xi’an Jiaotong University and is pursuing a Ph.D. degree at the MOE Key Laboratory for Intelligent Networks and Network Security, Xi’an Jiaotong University. His research interests include Internet traffic measurement and modeling, multimodal learning, anomaly detection, and botnet analysis.
\end{IEEEbiography}
\vspace{-0.5cm}

\begin{IEEEbiography}[{\includegraphics[width=1in,height=1.25in,clip,keepaspectratio]{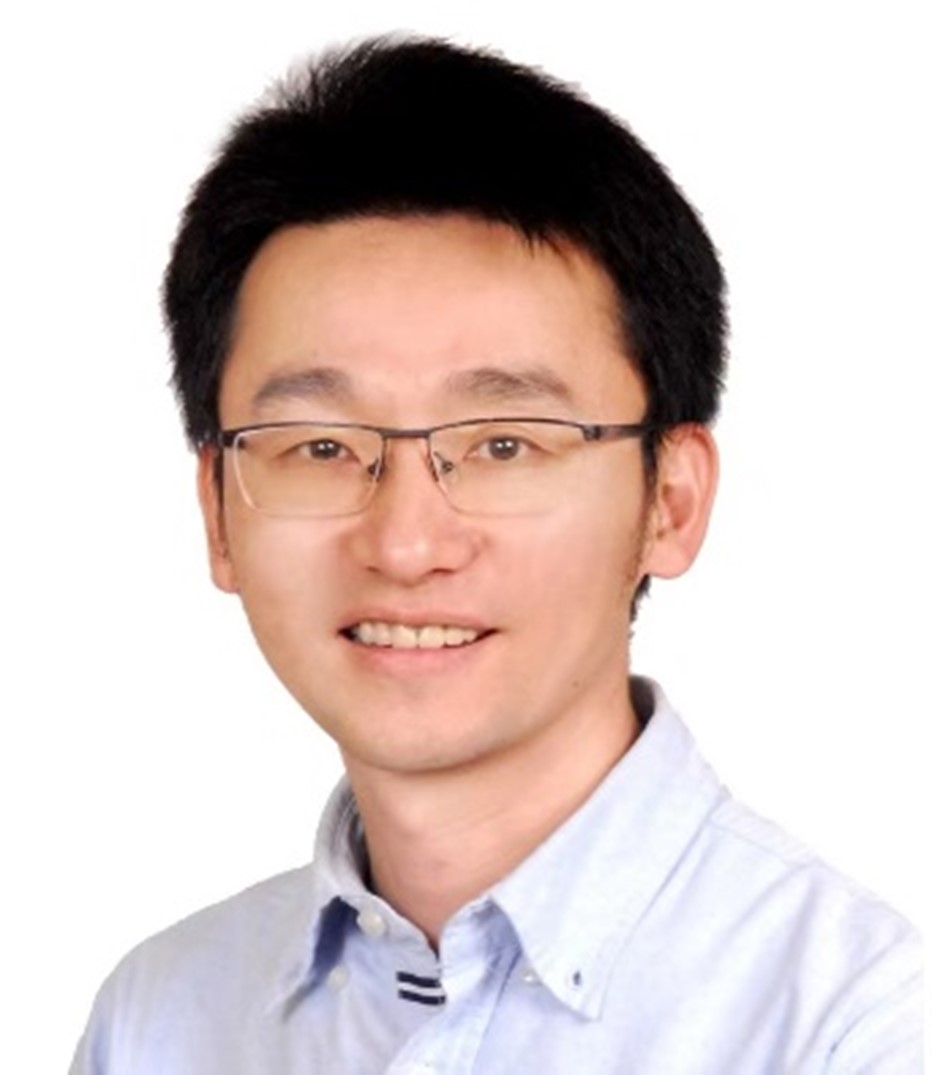}}]{Zhou Su}
has published technical papers, including top journals and top conferences, such as {\scshape IEEE Journal on Selected Areas in Communications}, {\scshape IEEE Transactions on Information Forensics and Security}, {\scshape IEEE Transactions on Dependable and Secure Computing}, {\scshape IEEE Transactions on Mobile Computing}, {\scshape IEEE/ACM Transactions on Networking}, and {\scshape INFOCOM}. Dr. Su received the Best Paper Award of International Conference IEEE ICC2020, IEEE BigdataSE2019, and IEEE CyberSciTech2017. He is an Associate Editor of {\scshape IEEE Internet of Things Journal}, {\scshape IEEE Open Journal of the Computer Society}, and {\scshape IET Communications}.
\end{IEEEbiography}

\vfill

\end{document}